\documentclass[10pt,twocolumn,twoside]{IEEEtran}
\def\hlFlag{no}  
\def\grayFlag{no} 
\def\commentFlag{no}  

\usepackage[caption=false,font=footnotesize]{subfig}

\ifCLASSINFOpdf
\usepackage[pdftex]{graphicx}
\else
\usepackage[dvips]{graphicx}
\fi

\usepackage{url}
\usepackage{algorithmic}
\usepackage{algorithm}
\usepackage{color}
\usepackage[cmex10]{amsmath}
\usepackage{amsfonts}
\usepackage{amssymb}
\usepackage{ifthen}
\usepackage{hyperref}
\usepackage{enumitem}
\usepackage{array}
\usepackage[numbers,sort]{natbib}

\usepackage[latin1]{inputenc}
\usepackage{tikz}
\usetikzlibrary{shapes,arrows}
\usepackage{varwidth}

\usepackage{lipsum}
\ifthenelse{\equal{\hlFlag}{yes}}{\usepackage{soul}\usepackage{ulem}}{}
\ifthenelse{\equal{\hlFlag}{no}}{\newcommand\hl[1]{#1}}{}
\ifthenelse{\equal{\hlFlag}{no}}{\newcommand\xout[1]{}}{}

\ifthenelse{\equal{\grayFlag}{yes}}{ \graphicspath{{./figs_g/}{./}} }{ \graphicspath{{./figs/}{./}} }

\makeatletter
\newcommand{\mathleft}{\@fleqntrue\@mathmargin\parindent}
\newcommand{\mathcenter}{\@fleqnfalse}
\makeatother

\newcommand\comment[1]{{\it \sffamily\tiny\textcolor{blue}{#1}}}
\ifthenelse{\equal{\commentFlag}{no}}{\renewcommand\comment[1]{}}{}

\begin{document}
\title{Unsupervised Incremental Learning and Prediction of \hl{Music} Signals}
\author{Ricard~Marxer, Hendrik~Purwins%
\thanks{R. Marxer is with 	Speech and Hearing Group, Department of Computer Science, University of Sheffield, 
Regent Court, 211 Portobello Street, Sheffield,
S1 4DP, UK, r.marxer@sheffield.ac.uk.}%
\thanks{H.Purwins is with Audio Analysis Lab and Sound and Music Computing Group, 
Aalborg Universitet  K{\o}benhavn,
A.C. Meyers V{\ae}nge 15, 2450 Copenhagen SV, hpu@create.aau.dk.} %
\thanks{R. Marxer and H. Purwins were with Music Technology Group, %
Universitat Pompeu Fabra, Roc Boronat, 138, %
08018 Barcelona, Spain.}%
\thanks{H. Purwins was also with Neurotechnology Group, %
Berlin Institute of Technology, %
Sekr. MAR 4-3, %
Marchstr. 23, 10587 Berlin, Germany.}
\thanks{This work was partially funded by the EmCAP (European Commission FP6-IST,
contract 013123) and INSPIRE-ITN (European Commission FP7-PEOPLE-2011-290000) projects. Thanks to Daniel Bartz from Berlin Institute of Technology for advice in statistics and for
proof reading.}%
}
\maketitle

\begin{abstract}
A system is presented that segments, clusters and predicts musical audio in an unsupervised manner, adjusting the number
of (timbre) clusters instantaneously to the audio input.
A sequence learning algorithm adapts its structure to a
dynamically changing clustering tree.
The flow of the system is as follows: 
1) segmentation by onset detection, 2) timbre representation of 
each segment by  Mel frequency cepstrum coefficients,
3) discretization by incremental clustering, yielding a tree of different sound classes (e.g. 
instruments) that can grow or shrink on the fly driven by the instantaneous sound events, resulting in a 
discrete symbol sequence,
4) extraction of statistical regularities of the symbol sequence, using hierarchical N-grams 
and  the newly introduced conceptual Boltzmann machine,
and 5) prediction of the next sound event in the sequence.
The system's robustness is assessed  with respect to complexity and noisiness of the 
signal. 
Clustering in isolation yields an adjusted Rand index (ARI) of   82.7\% / 85.7\% for data sets of singing voice and drums.  
Onset detection jointly with clustering achieve an ARI of 81.3\% / 76.3\%  and the prediction of the entire system yields an ARI of 27.2\% / 39.2\%. 
\end{abstract}


  \maketitle
\section{Introduction}
\comment{JASA allows maximum of 12 pages (if more high costs). Color pictures are charged extra.(make sure all pictures are readable in black and white)}
\comment{JASA style file adaptation http://scitation.aip.org/journals/doc/ASALIB-home/corp/pdf/jasa/author\_contrib.pdf,
Todo:
    *some issues:
      **discuss why starting with this particular parameter grid
      **check that scalar and index names are minimally ambiguous across all the levels of the method
    * discuss
          **manual acuity determination
          **what happens for Boltzmann expectation n-gram length 1?
          **To how many clusters do the acuity values refer to?
    *Give it to someone for proof read (Bob Sturm)
    }

\comment{The human music listener differs in a number of crucial aspects from a traditional computer system in analyzing musical content. The differences are in the learning paradigm, flexibility, and the role of expectation underlying the music analysis process.  We enjoy music even if we do not have a formal education in music theory. Implicitly, we understand musical structure even without being able to name a sound phenomenon such as a 5/4 meter, a dominant seventh chord or a {\it Fl{\"u}gelhorn}. On the other hand, traditional music information retrieval systems need to be taught by a large quantity of labeled training examples, to learn explicitly what a 5/4 meter, a dominant seventh chord, or a {\it Fl{\"u}gelhorn} is. 
Almost everyone can follow the brass section in a big band. And if we listen more carefully we can distinguish between a trombone, a trumpet, and a {\it Fl{\"u}gelhorn} solo.  On the other hand, a traditional music information retrieval system that is only trained to recognize a trumpet and a trombone will be challenged when confronted with a {\it Fl{\"u}gelhorn}. It does not have the flexibility to recognize the {\it Fl{\"u}gelhorn} as a novel musical instrument sound. Listening to Mozart, a dominant seventh chord invokes in us the expectation of hearing a tonic chord and a half-finished  {\it Fl{\"u}gelhorn}  melody will make us believe that the melody will be continued by the  {\it Fl{\"u}gelhorn}. Traditional context-free music information retrieval systems classify an event as a dominant seventh chord or a {\it Fl{\"u}gelhorn} without considering previous sounds.}

\comment{Daniel Bartz: Intro should more consider our system and the motivation for it. Solution (trial):  } Human music listening  adapts to novel acoustic stimuli  and is largely based on unsupervised learning, in contrast to most traditional music analysis systems. For music transcription \citep{downie05}, prediction \citep{conklin1995multiple,pearce2004}, representation \citep{mozer1994,lartillot2001}, automatic accompaniment, or human-machine-improvisation \citep{pachet2003,assayag2004}, a traditional system usually is based either on symbolic data instead of audio input,  or on classifiers that are pre-trained on a labelled data base \citep{downie05}. If a system, based on pre-trained classifiers needs to cope with new musical concepts (instruments, harmonies, pitches, motifs) it has not been designed for, it may cease to work reasonably. Such a system would have to be retrained with labeled data, every time a new instrument (pitch, harmony etc.) appears. This presents a severe lack of flexibility of such a system, in contrast to human cognition processing new instruments and harmonies with ease, even if one has not heard them before. A human mind can grasp a novel motif, when listening to a piece or an improvisation. Unsupervised learning (clustering) instead of supervised classification is one paradigm how an algorithm can model the cognition of novel concepts \citep{hazan2009,marx10,marchini2010a}. Based on a discrete representation of the input derived by  clustering, an n-gram, i.e.\ a suffix tree, can be used as a statistical representation of the structure of the input sequence \citep{conklin1995multiple,pearce2004}. In this paper, we extend such a system by equipping it with the capability to deal with a varying number of clusters. The number of clusters can increase if a new instrument appears. The cluster number decreases if two instruments become to sound very similar. We implement these features by using  unsupervised \emph{online learning}.  This requires that the n-gram (suffix tree) must be coupled with the clustering in order to be able to merge or split the symbol counts when cluster numbers change. We introduce a system prototype that learns in an unsupervised, adaptive manner and that generates predictions from audio sequences. From the first note it will begin to generate reasonable predictions without using previous knowledge. 

Many previous approaches to predicting musical sequences are based on symbolic representation  \citep{mozer1994,conklin1995multiple,lartillot2001,pachet2003,assayag2004,pearce2004}. 
\citet{paie2009} present a model that is capable of predicting and generating melodies using  a combination of  Bayesian networks, clustering, rhythmic self-similarity and a special representation of melody.\comment{The method exploits the self-similarity of a piece and the dyadic organization of its
rhythmic structure.} The distances between rhythmical patterns are clustered and the continuation of a melody  is predicted conditioned on the chord root, chord type, and Narmour group of recent melodic notes. 
\citet{hazan2009} build a system for generation of musical expectation that operates on music in audio data format. 
The  auditory front-end segments the musical stream and extracts both timbre and timing description. 
In an initial bootstrap phase, an unsupervised clustering process builds up and maintains a set of different
sound classes. The resulting sequence of symbols is then processed by a multi-scale technique based on n-grams.
Model selection is performed during a bootstrap phase via the  Akaike information criterion.
\citet{marchini2010a} present a non-adaptive system that learns rhythmic patterns from drum audio recordings and synthesizes music variations from the learned sequence. The procedure uses a fuzzy multilevel representation. Moreover, a tempo estimation procedure is used to guarantee that the metrical structure is preserved in the generated sequence.  
Online clustering has been proposed by \citet{zhan05} for document clustering. \citet{bertin2010clustering} have used online k-means to cluster beat-chroma patterns. The Hierarchical Dirichlet Process Hidden Markov Model (HDP-HMM)  \cite{teh_hierarchical_2010} has been used for segmentation in conjunction with clustering.  
\citet{fox_sticky_2011} and  \citet{ren_dynamic_2008} have proposed 'sticky' versions of the HDP-HMM that introduce explicit modelling of state occupancy duration. These models are applied to segmentation of a Beethoven sonata into musical sections  \cite{ren_dynamic_2008} and to speaker diarization \cite{fox_sticky_2011}. 
\citet{stepleton_block_2009} used the block diagonal infinite hidden Markov model
for musical theme labelling. However, these methods do not perform incremental online learning, whereas we propose an online incremental clustering method that uses a separate segmentation method (onset detection) and switches relatively rapidly between states. \citet{bargi2012} have adapted HDP-HMM to an online setting employing an initial supervised learning phase (bootstrap) whereas our approach  is entirely unsupervised.  

A part of the work covered in this paper, the application of the hierarchical n-grams on the \emph{Voice} data, has been presented previously \citep{marx10}. Here we compare that method with the conceptual Boltzmann machine and with HDP-HMM on an extended data set using a more advanced evaluation measure (the adjusted Rand index) and providing more examples of adaptive clustering. 
We will give an overview of the system, introduce its components, namely segmentation, timbre representation, clustering, and prediction.
Then we will introduce the adjusted Rand index, test the performance of the sequence analysis algorithms under noisy conditions, of each system module separately, and in conjunction. Finally, we will give some demonstration examples. Audio-visual data and examples are available
on the supporting website \citep{marxer14web}.

\section{System Overview}
\begin{figure}[htbp]
\begin{center}

\begin{tikzpicture}[%
scale=0.75,
>=stealth,
node distance=5em]
	\tikzstyle{every node}=[%
	align=center,
	font=\scriptsize,
	execute at begin node={\begin{varwidth}{5em}},
	execute at end node={\end{varwidth}}],
	text centered,
	inner sep=0pt]
	\tikzstyle{label} = [minimum height=2em]
	\tikzstyle{block} = [rectangle, draw, thick, rounded corners, minimum height=2em]
	\tikzstyle{line} = [draw, thick ,->]

    \node [block] (segment) {Segmentation};
    \node [label,left of=segment,xshift=1em] (input) {Audio};
    \node [block,right of=segment] (feature) {Feature\\Extraction};
    \node [block,right of=feature] (clustering) {Incremental\\Clustering};
    \node [block,right of=clustering] (sequence) {Sequence\\Analysis};
    \node [label,xshift=-1em,right of=sequence] (output) {Next\\Symbol\\\& IOI};
    \path [line] (input) edge[->] (segment);
    \path [line] (segment) edge[->] node[anchor=north,yshift=-1em] {Segments} (feature);
    \path [line] (feature) edge[->] node[anchor=north,yshift=-1em] {Timbre\\Representation} (clustering);
    \path [line] (clustering) edge[->] node[anchor=north,yshift=-1em] {Symbol\\Sequence} (sequence);
    \path [line] (sequence) edge[->] (output);

    \path [line] (feature) edge[->,bend left=45] node[anchor=south] {Growth/Shrinking\\of Clustering Tree} (clustering);
    \path [line] (clustering) edge[->,bend left=45] node[anchor=south] {Structural\\Adaptation} (sequence);
\end{tikzpicture}
\end{center}
\caption{System architecture: An audio sound file 
is segmented, using onset detection. Each  segment  is then represented as a high-dimensional timbre feature vector which is 
clustered into symbols. 
Symbols are added or removed to the clustering tree on the fly.  The symbol sequence is then statistically analysed, adapting to the varying number of symbols, allowing for prediction of the next symbol and the next inter-onset interval (IOI).}
\label{fig:generalarch}
\end{figure}
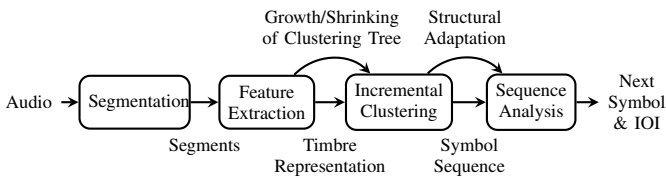

The system that we present in this paper (cf.\ Fig.~\ref{fig:generalarch})  consists of four main stages: segmentation  by onset detection, feature extraction resulting in timbre representation, incremental clustering giving a symbol sequence, and  sequence analysis yielding a prediction of the next symbol. In particular, the clustering tree generated by incremental clustering grows and shrinks online, driven by the most recent sounds. In turn, the sequence model adapts to the changing numbers of symbols. Segmentation and representation can be interpreted as a model of  perception, whereas discretization and prediction can be considered to be a cognitive model.

\subsection{Segmentation by Onset Detection}
\label{sec:preprocessing}
In this section, we will explain how to segment an audio stream into events, using onset detection. 
In order to be more generally applicable, we have employed the complex domain based onset detector \citep{duxb03}, since it subsumes onset detection algorithms based on energy, spectral difference, or phase as special cases.
This onset detection function captures onsets due to abrupt energy changes 
as well as soft onsets induced by  pitch changes, with little energy variations. 
For each frame $l$, the short-term Fourier transform yields a complex spectrum
$ X_{k}(l) = r_{k}(l) e^{i \phi_{k}(l)},$
with magnitude $r_{k}$ and phase  $\phi_{k}$ for the $k$-th bin with frame length $K$  $ (0\le k \le K-1)$. 
We build the onset detection function as the Euclidean distance between the actual complex spectrum   $ X_{k}(l)$ at bin $k$ and the estimated complex spectrum \citep{duxb03}:%
\begin{equation}
 \hat{X}_{k}(l) = \hat{r}_{k}(l) e^{i \hat{\phi}_{k}(l)},
\end{equation}
where the estimated amplitude $\hat{r}_{k}(l)$ is set equal to the magnitude
of the previous frame $\left \| X_{k}(l-1)\right \|$, and the estimated phase 
$\hat{\phi}_{k}(l)$ is calculated as the linear extrapolation from the unwrapped phases of the two preceding frames:
\begin{equation*}
 \hat{\phi}_{k}(l) = \mbox{princarg} \left [ \tilde{\varphi}_{k}(l - 1) +(\tilde{\varphi}_{k}(l - 1)  - \tilde{\varphi}_{k}(l - 2))\right ],
\end{equation*}
where the $\tilde{\varphi}$ denotes the unwrapped phase  
and the princarg operator maps the unwrapped value back to the $(-\pi,\pi]$ range.
We calculate the bin-wise Euclidean distance between the actual and the estimated complex spectrum, quantifying the stationarity for the $k$-th bin as: 
$ \Delta_{k}(l) = \| X_{k}(l) - \hat{X}_{k}(l) \|.$
By summing across all $K$ bins and across $M+1$ consecutive frames centered around frame $l$ (smoothing), we yield the onset detection function:
\begin{equation}
 \eta(l) = \frac{1}{M}\sum_{j=\lceil \frac {-M}2\rceil}^{\lfloor \frac M2 \rfloor} \sum_{k=0}^{K-1} \Delta_{k}(l+j).
\label{eqn:odfsmoothing}
\end{equation}
Similarly to previous approaches \citep{bell03},  
an adaptive  threshold $\theta(l)$ is used. This threshold  is calculated as the scaled median across a look-ahead
window of length $P +1$
\begin{equation}
  \theta(p) = C \cdot \text{median}_{n \in (p, p+1,\ldots, l + P)}(\eta(n)),
\label{eqn:odfthreshold}
\end{equation}
with $0\le C\le 1$ being a predefined parameter  controlling 
the sensitivity of the onset detector.  
\comment{to prevent confusion with the categories C, another variable name, not C?}
In order to eliminate multiple occurrences of onsets shortly one after another, smoothing is applied via
another window of length $W+1$ centered at sample $l$: 
\begin{equation}
  \mu(l) = \sum_{m=\lceil -\frac W2 \rceil }^{\lceil \frac W2 \rceil}  \max(\eta(l+m)-   \theta(l+m),0)
  \label{eqn:max_window}
\end{equation}
A silence threshold $\theta_s$ is applied:
\begin{eqnarray}
  \mu_{s}(l)  = \max(\mu(l)-   \theta_s,0).
 \label{eqn:silence_threshold}
\end{eqnarray}
 Finally, the  local maxima of $\mu_{s}(l)$ define the predicted onset times.  
     
\subsection{Feature Extraction for Timbre Representation}\label{sec:timbre_rep}
For each onset, a short window of length $L$ subsequent to the onset time  is analyzed.
For each frame within this window, the first 13 Mel-Frequency Cepstrum Coefficients (MFCC) \citep{mermelstein1976}  are calculated.
To model the coefficient's temporal behaviour right after the onset, 
for each coefficient another Discrete Cosine Transform (DCT) is calculated on the sequence of coefficients across the 
frames. Taking the first 4 DCT coefficients for each MFCC yields a 52-dimensional vector, representing  timbral 
features both of the sound event's spectral characteristics and their initial temporal development.

\subsection{Incremental Clustering for Symbol Sequence Generation}
\label{sec:clustering}
The clustering stage receives multivariate feature vectors from the preprocessing
stage and converts them into symbols. 
It is important to state that in our system the events are clustered in an online manner and in order of arrival, since this symbolic representation is used immediately to create predictions of future events. As a reference and benchmark, we compare  online clustering by Cobweb with a  state-of-the-art batch clustering method exploiting sequential information, the HDP-HMM.
\subsubsection{Cobweb}
For this purpose, \citet{marx07} used the Cobweb \citep{fish87}. Cobweb is an incremental clustering model 
which continuously builds a knowledge tree (hierarchical partitioning of the object space) and assigns to each instance
 a partition created at each level until the object reaches the leaves of the tree.
 Each node of the tree represents a concept.  A concept is modelled by a univariate Gaussian for each feature dimension.  
The edges of the structure represent taxonomic relations.  Further works \citep{mcku90,yoo95} have proposed techniques to create, 
in an unsupervised manner, the concept tree based on the sequence of data presented, by the use of a heuristic function to be maximized.
The heuristic function used in this paper is the numerical version of the standard category utility function used by Fisher 
and introduced by \citet{gluc85}. The version of Cobweb that we will use was presented as Cobweb/3 \citep{mcku90} and
later extended as Cobweb/95 \citep{yoo95}. 
This algorithm clusters  $D$-dimensional feature vectors ${\mathbf x}=(x_1,\ldots, x_D)$ extracted in the previous section.
Consider a particular cluster containing  $I$ feature vectors.  Let $\sigma_d$ be the standard deviation in component $d$ of the input feature vectors
assigned to that cluster. Then $\sum^D_{d=1} \frac 1{\sigma_d}$ is the specificity of that cluster across all feature dimensions.
We consider the utility $U$ to quantify the gain in specificity by splitting this cluster into $K$ child clusters.  For a potential child cluster  $1 \le k \le K$   with $I_k$ instances and each input feature dimension $d$ we define $\sigma_{dk}$ to be the inner cluster standard deviation in that dimension.
Then $\sum_{d=1}^D \frac 1{\sigma_{dk}}$ is the specificity of cluster $k$, and   $\sum^K_{k=1} \frac {I_k}I \sum^D_{d=1} \frac 1{\sigma_{dk}}$ is the specificity of the child clusters altogether.
 For the cluster utility holds

\begin{equation}
  U \propto  \frac 1K \left(\sum^K_{k=1} \frac {I_k}I  \sum^D_{d=1}{\frac{1}{\max(\sigma_{dk}, a)}} - \sum^D_{d=1}{\frac{1}{\sigma_d}} \right),
\label{eqn:cobweb}
\end{equation}

\comment{can't find the equation. The report isn't online anymore.}
The acuity parameter $a$ is an upper limit of maximal specificity (minimal standard deviation) of the clusters, thereby controlling the maximal resolution of the clustering discrimination.

The incorporation of an object is a process of clustering the object by descending the tree along an appropriate path, 
updating counts along the way, and possibly performing one of several operations at each level. These operators are:
\begin{itemize}
\item creating a new node,
\item removing all children from a node (pruning),
\item combining two clusters into a single node, and
\item splitting a node into several nodes.
\end{itemize}
While these operations are applied to a single object set partition (i.e., set of siblings in the tree), 
compositions of these primitive operations transform a single clustering tree. As a search strategy 
we use hill-climbing through a space of clustering trees.

Thereby, the input is converted into a sequence not only of symbols, but also of meta symbols (partitions)
according to their parent nodes and grandparent nodes in the cobweb tree.
The symbols and meta symbols provide the alphabet on which expectations
will be generated by the hierarchical N-gram.

We modify the set of possible Cobweb operations (see above) in order to achieve persistent
partitioning.
This reduced set of operations can perform any of Cobweb's original operations.  
We reformulate the second Cobweb operation (see above) in order to control the clustering only by new incoming events.
Other partitions and past events should not be considered. 
This reduces the operations to:
\begin{itemize}
 \item creating a new partition inside a \textit{container partition}, 
 \item removing a partition, reparenting it's children if it has any.
\end{itemize}

\subsubsection{Hierarchical Dirichlet Process Hidden Markov Model (HDP-HMM)}
The feature vector sequence may also be modelled as the emission of a HDP-HMM, a Bayesian nonparametric model in which the hidden states can be considered as clusters. Given the observed feature vector sequence, the most likely hidden state sequence can be interpreted as a sequence of symbols.  In the HDP-HMM, the hidden states are assumed to be drawn from a countably infinite state space. The HDP-HMM is used to jointly estimate the number of clusters, the cluster assignment of the feature vectors, and the transition
probabilities between clusters. Inference in the HDP-HMM is performed using the weak limit approximation \citep{fox_sticky_2011} implemented in \texttt{pyhsmm} \citep{johnson2013hdphsmm}.\footnote{http://github.com/mattjj/pyhsmm}
However the inference does not work in an online manner, it requires the entire feature vector  sequence as input. This method is offline (batch mode) and is only used as a reference and benchmark, since it does not fulfil the constraint of clustering the feature vectors as they arrive to perform immediate prediction from the very beginning.

\subsection{Sequence Analysis for Next Symbol/Onset Prediction}
\label{sec:expectation} 
We choose two methods (hierarchical N-grams and conceptual Boltzmann machine) \citep{pfle02} that require relatively little storage by deducing frequency counts for longer sequences from frequency counts of their shorter subsequences. These algorithms iteratively predict the next symbol (or the inter-onset interval= IOI respectively) $c_{t+1}$  based on previous symbols (IOIs) $c_{t-n+1},c_{t-n+2},\ldots, c_{t},$ previously generated by incremental clustering. Thereby we derive which sound to expect when.
The prediction of symbols and of IOIs is performed independently. By predicting the IOI, we can determine 
the onset time of symbol $c_{t+1}.$
\subsubsection{Hierarchical N-Grams (HN)}
$N$-grams have been used in the analysis of genome sequences and in language modeling \citep{zitouni2002backoff}. Exhaustive $N$-grams count the instances of all possible symbol (IOI) sequences of length $N.$ Their memory requirement is exponential in the sequence length $N$ and the problem arises how to account
for patterns that have not occurred before (zero frequency problem). We use $N$-grams as an estimate for the forward conditional distribution for online prediction of the next symbol (IOI). 

Hierarchical $N$-grams (HN) \citep{pfle02} need less memory than exhaustive $N$-grams.
HN are a combination of sparse $N$-gram models in a hierarchical structure that allows compositional learning.  Compositional learning consists in learning long patterns from already learned sub-patterns. 
In sparse $N$-grams counts of the most frequent patterns and a separate total count for the non-frequent patterns are kept.  This technique separates the estimates of patterns whose statistics are reliable from the estimates of infrequent patterns whose statistics are biased.
On the other hand, the multi-width exhaustive approach consists in keeping the count of all possible patterns of at most length $N$.  These models are able to represent any distribution of patterns up to width $N$.

Let  ${\cal C}_1=\{c^1,\ldots,c^{|{\cal C}_1|} \}$ be the set of cluster indices, renumbered so that they reflect the order of their first appearance in the symbol sequence $\mathbf{c}=(c_1,\ldots, c_t),$ achieved from the  clustering process in the previous section.
 ${\cal C}_1$ forms the alphabet of the n-gram. 
 Then,  ${\cal C}_n$ is the set of all possible $n$-grams of length $n$ composed from alphabet  ${\cal C}_1$. To exploit sparsity, we only consider the patterns that have actually occurred as a subsequence of  $\mathbf{c}$ so far until time $t$. The  set  of patterns of length $n$ having occurred so far will be denoted by ${\cal C}_n=\{\mathbf{c}^1,\ldots \mathbf{c}^{|{\cal C}_n|}\}$,  in which again  the subpatterns  are ordered  according to their first appearance.  $o(\mathbf{c})$ is defined as the position of  $\mathbf{c}$ in ${\cal C}_{|\mathbf{c}|}.$
We consider HNs of maximal length $N$. 
Let  $C_{n, i} (n\le N)$ be the frequency count of the $i$-th pattern of length $n$ and let $T_{n, i}$  be the  total count of patterns
of length $n$ since pattern $i$ occurred for the first time. In Algorithm~\ref{algo:ngram}, we use the counts $T_{n, i}$ and $C_{n, i}$ to iteratively  estimate the joint probabilities $P_{n,i}$ for all patterns
seen so far. We define $T_{n,0}:=T_{1,1}$ for $1 \le n \le N$.
In simple $N$-grams, the empirical frequency  $\frac {C_{n,i}}{T_{n,1}}$ could be used as an estimate for the probability  of a pattern of length $n$.
In the HN method  (Eq.~\ref{eqn:p}), the  probability $P^{n-1}_{n,i}$ for the $i$-th pattern of length $n$ under the joint distribution of width $n-1$ is estimated. For pattern $\mathbf{c}^i=(c^i_1,\ldots,c^i_n),$ statistical estimates (Eq.\ 8.2 in \citeauthor{pfle02})
\begin{eqnarray}
P^{n-1}_{n,i} = \frac {P(c^i_1,\ldots,c^i_{n-1}) \cdot P(c^i_2,\ldots,c^i_n)}{\sum_{y\in {\cal C}_1} P(c^i_2,\ldots,c^i_{n-1},y)}
\label{eqn:width-1}
\end{eqnarray}


are calculated, using  the sub-patterns of the $i$th pattern of lengths  $n$. We estimate the probability $Q^{n-1}_{n,i}$ (Eq.~\ref{eqn:q}) of sub-patterns of length  $n-1$ of the $i$th pattern of length $n$ of not being a subpattern of the first $i$  patterns of length $n$. %
In Eq.~\ref{eqn:p}, they are  weighted by their confidence. 
 The confidence values depend on the number of occurrences of the patterns.  Therefore, when a pattern of length $n$ has appeared rarely in the data stream, its probability of occurrence is estimated from a small number of counts and is not reliable. 
In this case the probability of appearance is better estimated from the $n-1$ length sub-patterns through $P^{n-1}_{n,i}$. 
In other words, the information of patterns of large lengths is integrated with the information of models of small lengths.
Pfleger shows that the probability of a given pattern can be calculated in a linear sweep by updating all the probabilities in order of the pattern's first occurrence and length.

\begin{algorithm*}
\caption{The Hierarchical N-Gram for Merged Clusters}
\label{algo:ngram}
\begin{algorithmic}
\STATE Initialization ${\cal C}_n=\{\} $ for $1 \le n \le N$
\FOR { incoming event $c_t$} 
\FOR {$ 1\le n \le N$}
\IF { $(c_{t-n+1},\ldots,c_t) \notin {\cal C}_n$}   
\STATE  Add new pattern: ${\cal C}_n={\cal C}_n \cup (c_{t-n+1},\ldots,c_t),$ $T_{n, |{\cal C}_n|}=0, C_{n, {|{\cal C}_n|}}=0$  
\ENDIF
\IF {$\mathbf{c}^1,\ldots, \mathbf{c}^k\in  {\cal C}_n$ are merged by Cobweb}
\STATE
\vspace{-1em}
\mathleft 
\begin{flalign}\label{eqn:mergeset} 
o'&=\min (o(\mathbf{c}^1),\ldots, o(\mathbf{\mathbf{c}^k})) \\
C_{n, o' }&=\sum_{i=1}^k C_{n, o(\mathbf{c}^k) }  \label{eqn:mergesum} \\
{\cal C}_n  &={\cal C}_n\backslash \{ \mathbf{c}^1,\ldots, \mathbf{c}^{o'-1}, \mathbf{c}^{o'+1}, \ldots,   \mathbf{c}^{k}\} 
\label{eqn:mergeremove}
 \end{flalign}  
\mathcenter
\vspace{-1em}
 \STATE Update indices
\ENDIF
\STATE  Update counts: $C_{n, o(c_{t-n+1},\ldots,c_t)}=C_{n, o(c_{t-n+1},\ldots,c_t)}+1$
\STATE Update total counts: $T_{n, i}=T_{n, i}+1$ for $ 1 \le i \le | {\cal C}_n| $
\ENDFOR
\STATE Calculate joint probabilities:
\STATE $P^0_{1,i}= \frac 1{|{\cal C}_1 |} \;\; (1 \le i \le |{\cal C}_1 |)$ 
\FOR {$ 1\le n \le N$} 
\FOR {$1 \le i \le |{\cal C}_n |$}
\STATE
\vspace{-1em}
\mathleft
\begin{flalign}\label{eqn:q}%
Q^{m}_{n,i} &= (1-\sum_{k=1}^i P^{m}_{n,k}) \;\;\;\; (m=n,n-1) 
\end{flalign}
\mathcenter
\STATE \hspace{0.5cm} Calculate $P_{n,i}^{n-1}$ according to Eq.\ref{eqn:width-1}
\mathleft
\begin{flalign}
P^n_{n,i} &= \frac{1}{T_{1,1}} \left[ C_{n, i} + \sum_{j=0}^{i-1} (T_{n, j} - T_{n, j+1}) \cdot Q^n_{n,j}  \cdot \frac{P^{n-1}_{n,i}}{Q^{n-1}_{n,j}} \right]  \label{eqn:p}
\end{flalign} 
\mathcenter
\vspace{-1em}
\ENDFOR
\ENDFOR 
\ENDFOR
\end{algorithmic}
\end{algorithm*}

\begin{figure}
 \centering
 \includegraphics[keepaspectratio=true, width=8.0cm]{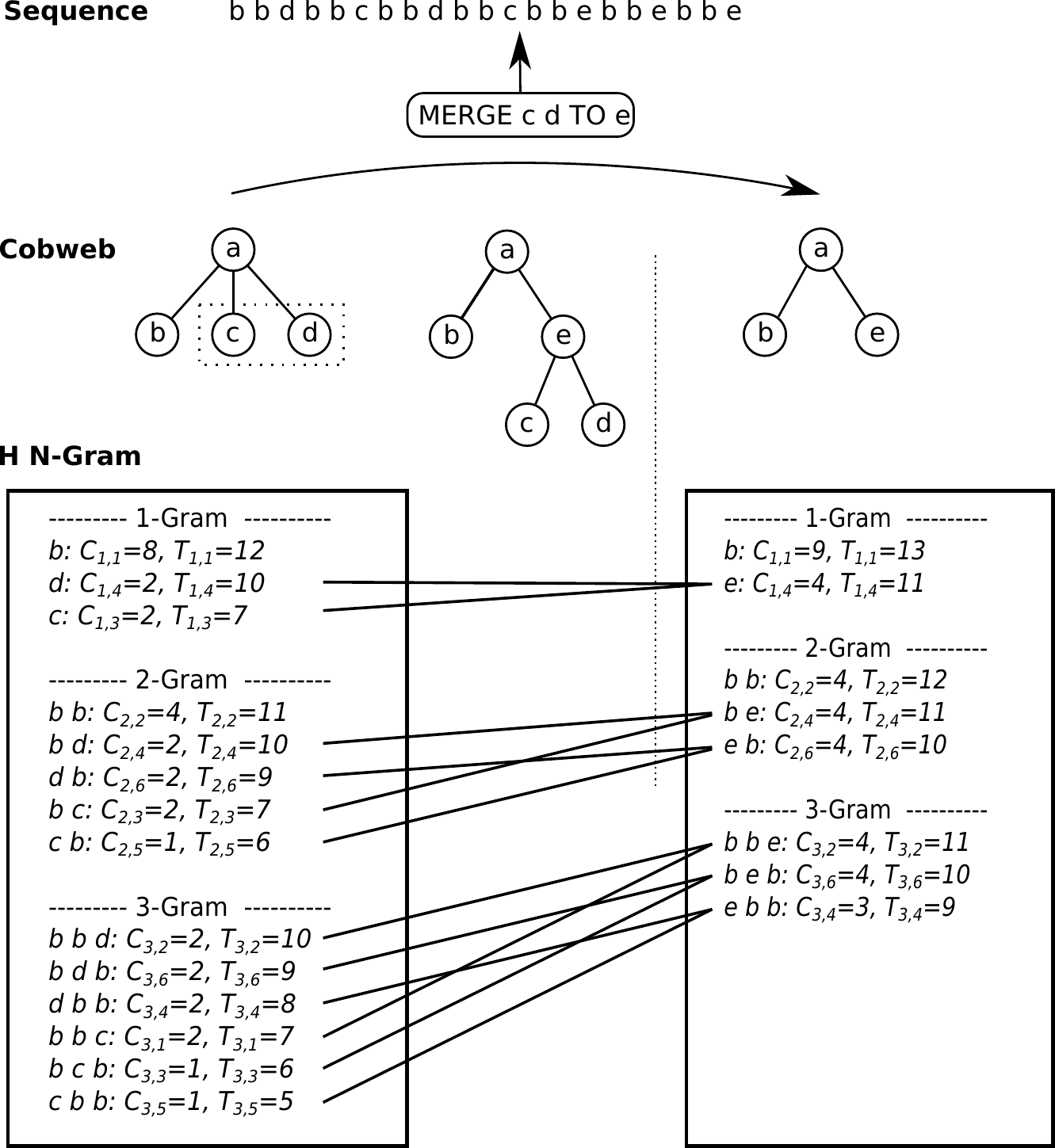}
 \caption{The Effect of a concept merge in the hierarchical n-gram. Nodes  \textit{c} and  \textit{d}  are merged into the new symbol  \textit{e}.  The $n$-gram inherits the counts for patterns including \textit{c} and  \textit{d}  to patterns including  \textit{e}.}
 \label{fig:conceptMergeHngram}
\end{figure}

In order to adapt Pfleger's HN \citep{pfle02} to our architecture, we have to link the operations of the clustering model to the operations on the $n$-gram (Fig.~\ref{fig:conceptMergeHngram}). When two or more clusters are merged in the clustering model, we have to remove the superfluous clusters from  the set of cluster indices (Eq.~\ref{eqn:mergeset})  and to sum up the counts for the merged clusters (Eq.~\ref{eqn:mergesum}).
For example, if the $n$-gram tracks patterns \textit{bbc} and \textit{bbd} and suddenly the clustering model merges symbols  \textit{c} and  \textit{d} into a new symbol  \textit{e}, the $n$-gram must sum up the counts of \textit{bbc} and \textit{bbd}  and substitute them with the count of \textit{bbe}.

\begin{figure}
 \centering
 \includegraphics[keepaspectratio=true, width=0.28\textwidth]{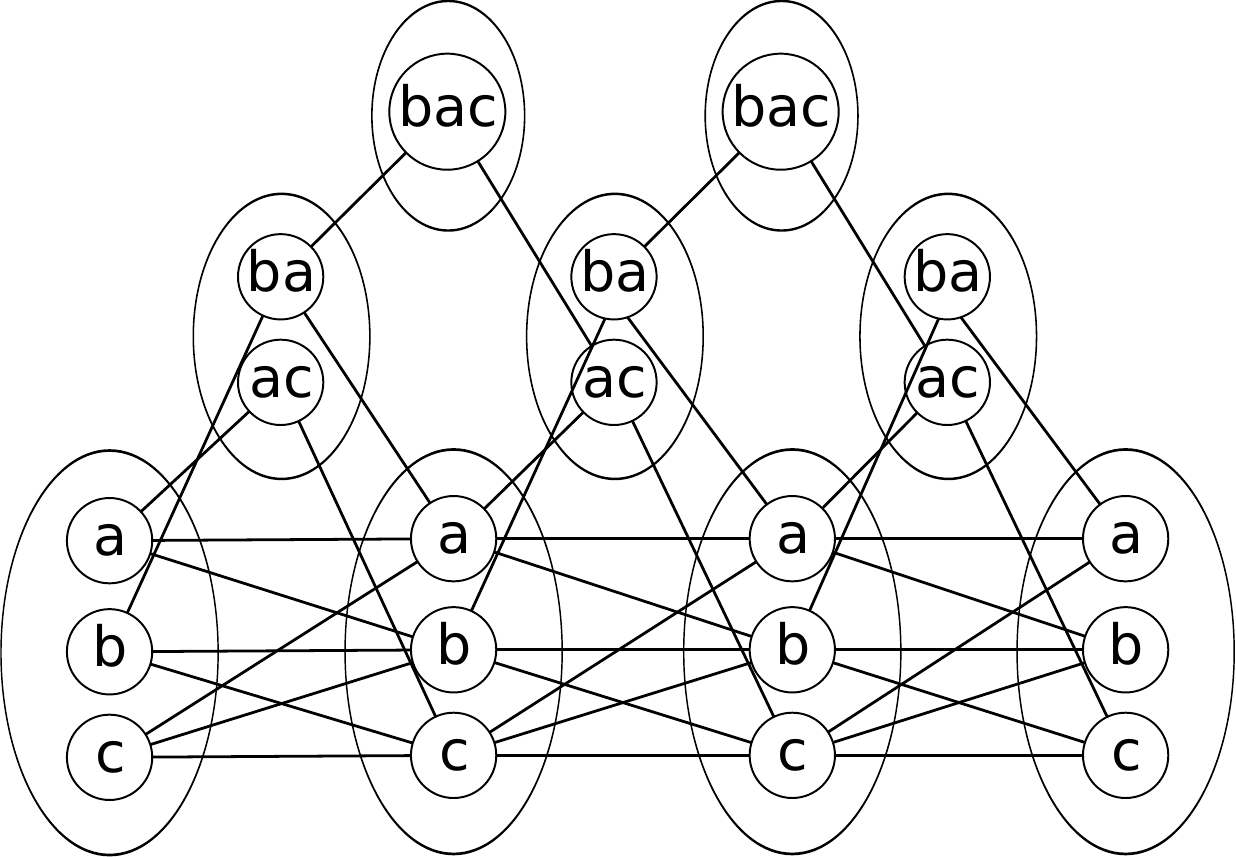}
\caption{Illustration of the continuous composition of symbols (atoms) in the Boltzmann Machine into longer patterns (chunks).}
\label{fig:catbm}
\end{figure}
\subsubsection{Conceptual Boltzmann Machine (CB)}

The Boltzmann machine \citep{ackley1985learning}  is a stochastic, symmetric-recurrent neural network that can be used to represent a joint distribution of random variables, to complete patterns, and in particular  (as in our case) to predict the continuation of a time series.

Formally, a Boltzmann machine consists of a vector of binary units  $(s_1,\ldots, s_I) \in \{0,1 \}^I $,   and symmetric weights $w_{ij} \in  {\mathbb R} $ between pairs of units $(s_i, s_j),$ an update rule for the units and 
a learning rule for the weights.   

Applied to \emph{categorical} data \citep{pfleger02}, a $V$-valued symbol $c_u$ is encoded as  binary units $s_{u_1},\ldots,s_{u_V}$ with $s_{u_v}=1$ if and only if $c_u=v$. To connect two $V$-valued variables $c_u$ and $c_i$, $V^2 $ weights $w_{i_j, u_v}$ are needed to connect the binary units representing the two variables.  Initially, the architecture of our particular Boltzmann machine implementation consists of sets of binary variables for consecutive symbols, where the binary nodes of each variable are initially only connected to the binary nodes of the previous and the next symbol. Depending on the other units and weights, the stochastic softmax update rule for the symbol is:
\begin{eqnarray}
P(c_i=j)=\frac 1{1+e^{-    \frac{ \sum_{u\not=i} \sum_{v=1}^{V}  s_{u_v} w_{i_j,u_v}}T}},
\label{eqn:unit_update}
\end{eqnarray}
with temperature $T$ decreasing from $T=50$ to $T=0.005$ in $100$ steps.
As an example of Gibbs sampling,  this  update rule is applied iteratively. In general, through \emph{simulated annealing} of the temperature $T$, the states converge to a particular state vector \citep{ackley1985learning}.
 
 For training the Boltzmann machine, the weights $w_{ij}$ have to be learned. 
 As in the case  of the restricted Boltzmann machine \citep{smolensky1986information}, in our case, not all pairs $(s_i,s_j)$ are connected by non-zero weights $w_{ij}$. 
 Units representing the same symbol are not connected among each other.
 For each binary previous symbol sequence, the update rule  (\ref{eqn:unit_update}) is iteratively  applied until  the final states are  reached (denoted by   $s^+_i$).  In addition, the update rule is applied with no units fixed until another vector of final states  ($s^-_i$) is reached.
 Then a stochastic gradient-based  learning step for the weights  can be performed with learning rate $\mu$ for a single training instance yielding  $s_i^+ s_j^+$:
\begin{eqnarray}
\Delta w_{ij}= \mu (s_i^+ s_j^+ - s_i^- s_j^- ).
\label{eqn:contrast_div}
\end{eqnarray}    
The learning step aims at minimizing the difference between $s_i^+ s_j^+$  and $s_i^- s_j^-.$ $\mu=0.1 $ is used.

When weight $w_{ij}$  rises above a threshold  $\theta_w$, a new \emph{hidden} unit is created, representing the concatenation of symbols connected by strong weights (cf.\ Fig. \ref{fig:catbm}) \citep{pfleger02}.  In addition,  weight $w_{ij}$ is removed. Iteratively, hidden units for   patterns of length $n+1$ are created from nodes representing patterns of  length $n$ and a new set of binary nodes representing patterns of length $n$ is appended. We set $\theta_w =0.2, 0.15, 0.1, 0.05$  respectively, depending on the length of the pattern the unit represents (length 1,2,3,4). This variant of the Boltzmann machine is called the \emph{compositionally-constructive categorical Boltzmann machine} \citep{pfleger02}.  For predicting the next symbol $c_{t+1}$, in a   trained  Boltzmann machine,
the respective  units are fixed to the previous symbol sequence $c_{t-n+1},\ldots, c_{t}$. After running the unit update rule (\ref{eqn:unit_update}) until convergence, the predicted next symbol  $c_{t+1}$  in the sequence can be retrieved from
the corresponding binary units of the Boltzmann machine.

In our system we have implemented a new method called \emph{conceptual Boltzmann machine (CB)}. In \citeauthor{pfleger02}, the Boltzmann machine acts on a static set of categories.  We have extended this to an architecture which operates on a dynamically changing taxonomy of categories.  Therefore, the model adjusts to the tree structure generated by the Cobweb.  This means the Boltzmann machine changes the architecture on the fly guided by the creation, removal, splitting, and merging operations suggested by the  Cobweb.  Accordingly, in the  Boltzmann machine,  the units and the update rule must be adjusted to the new structure.

During the run, sequences of atoms cause the creation of  higher-level chunks that represent patterns. The newly created chunks that represent patterns are then further chunked into nodes that represent patterns of longer length.  The longest pattern represented by a node is fixed to a value of $N$.

\comment{Please check parameters (especially the 4 weight chunking values) and equations. I inferred them from: conceptualboltzmann.py
Vertical weights: weights to hidden units?}

\section{Performance Analysis of the System}
\subsection{Measures for Clustering Evaluation}
\label{sec:clustereval}
Unlike in supervised learning, where accuracy can be measured between the annotated labels and the 
labels predicted by a classifier, the number of clusters predicted by the analysis can be different from the
number of annotated label categories. In addition, the mapping between annotated and predicted labels is unclear.
This creates the need for a particular clustering evaluation measure. The following measures for evaluating the agreement between annotations and predicted labels have been suggested:
purity \citep{zhao2005hierarchical}, F-measure \citep{larsen1999fast}, and Pearson's chi squared coefficient \citep{wagner2007comparing}, 
and Rand index \citep{wagner2007comparing}. We choose the latter measure for evaluation, since it is a natural extension of classifying elements to pairs of elements.  


A partition (clustering)  ${\cal C} $ of a set $\cal X$ is defined as a set ${\cal C}=\{{\cal C}_1,\ldots, {\cal C}_J \}$  of subsets ${\cal C}_j \subset {\cal X},$ so that $\cup_j {\cal C}_j  ={\cal X}$ and ${\cal C}_j,  {\cal C}_{j' } $   disjoint for $j\not=j'.$
Let $\mathbb{P}$ be the set of all partitions  of $\cal X$ and let $|{\cal C}|$ be the number of elements in a partition ${\cal C} \in \mathbb{P}.$
  Let ${\cal A}\in \mathbb{P}$ be a partition generated by annotation and let ${\cal C} \in \mathbb{P}$ be a predicted partition derived from an algorithm.
Let $\mathcal{P}=\{(x,x') | x,x'\in \mathcal{X}, x\neq x'\}$ be the set of pairs of distinct events. Let $\mathcal{L} \subset\mathcal{P}$
be the set of events pairs where both $x$ and $x'$ share  the same labels/annotation provided by $\cal A$ and let
$\mathcal{K} \subset\mathcal{P}$ 
be the set of event pairs where both $x$ and $x'$ lie in the same cluster provided by $\cal C$ . Then $| \mathcal{K} \cap \mathcal{L} |$ are the
number of point pairs that lie in the same cluster and - at the same time - share the same annotated  labels.
For $|{\cal C}|>1,$ the Rand index \citep{wagner2007comparing} is defined as:
    \begin{eqnarray}
    R({\cal A},{\cal C})= \frac {2 (|{\cal L} \cap {\cal K} | + |{\cal P}\backslash {\cal L}  \cap{\cal P}\backslash {\cal K}|)}{|{\cal C}| (|{\cal C}|-1)}.  
  \end{eqnarray}
Since  $R$ depends on the number of clusters $|{\cal C}|,$ we adjust the Rand index, comparing it with the expected
value of $R$ (baseline of a random clustering) $E_R$. 
The expected value of $R$ over all partition combinations $\mathcal{P} \times \mathcal{P}$ is  
calculated as \citep{fred2005combining}:
\begin{eqnarray} E_R = \frac 1{|{\cal C}|^2}
 \sum_{{\cal A,C} \in \mathbb{P}}  R({\cal A},{\cal C}).
\end{eqnarray} 
$E_R$ gets maximal for  ${\cal A}={\cal C}$:
 \begin{eqnarray} R_{\max} = \frac 1{| {\cal C} |^2}
 \sum_{{\cal C} \in {\cal P}}  R({\cal C},{\cal C}).
\end{eqnarray}
Then the \emph{adjusted Rand index (ARI)} holds:
 \begin{eqnarray} ARI({\cal A},{\cal C}) = \frac{R({\cal A},{\cal C}) -E_R}{R_{max}-E_R}
 \label{eqn:ari}
\end{eqnarray}
\comment{The problem is that the cluster measure varies for different clusters and for various numbers of data points.}
The ARI has values between $0$ (random partitioning) and $1$ (${\cal A} = {\cal C}$). 
\comment{What is the adjusted Rand score?}

The ARI assumes that annotations and clusterings are drawn randomly with a fixed number of clusters and a fixed number of elements per cluster \citep{wagner2007comparing}. Although this assumption will not always be true in our evaluation, we will use ARI, since it is a more established measure than alternative ones, such as the Fowlkes-Mallows index, the Mirkin metric, the Jaccard index \citep{meila_comparing_2007}, or entropy-based measures \citep{zhao2005hierarchical,wagner2007comparing}, e.g.\ normalized mutual information and variation of information.

In evaluating our system, we use the ARI in two ways: in the evaluation of 1) the clustering of the feature vectors of each event (Tables~\ref{tab:clustFrecall} and \ref{tab:clustFmeasureENST}) and of 2) the prediction of the entire symbol sequence, as explained in the sequel.
According to Fig.~\ref{fig:generalarch}, by segmentation, feature extraction, and clustering, the input
sound wave is transformed into a sequence $\mathbf{c}=(c_1,c_2, \ldots, c_T)$ of $T$ events, each one represented as  one of $J$ symbols. All occurrences of symbol $j$ can be included in a cluster ${\cal C}_j$ that contains 
all the indices $t$ where event $c_t$ equals symbol $j$. Then ${\cal C}=({\cal C}_1,{\cal C}_2,\ldots,{\cal C}_J)$ is a partition of ${\cal X}=(1,2,\ldots, T).$  To evaluate the prediction $\mathbf{c},$  we annotate one of
the ground truth labels $(1,2,\ldots, I)$ to each segment of the input, yielding an annotated sequence $\mathbf{a}=(a_1,a_2, \ldots, a_T).$ From this, a partition ${\cal A}=({\cal A}_1,\ldots,{\cal A}_I)$ can be generated in the same way as the partition ${\cal C}$ for $\mathbf{c}.$ The number $I$ of the annotated labels is not necessarily the same as the number of symbols $J$ determined by the clustering stage of our system. Then the ARI can be used to compare ${\cal C}$ and ${\cal A},$ as  done in Tables~\ref{tab:expecVoiceENST}-\ref{tab:predVoice} and Fig.~\ref{fig:patternLengthRepetitions}-\ref{fig:patternLengthSwitch}.

\subsection{Data Sets}\label{sec:data_sets}
Two sets of test data are employed: 
\begin{itemize}
\item Repetitive symbol sequences: We generate sequences that consist of patterns of length $n_l=2,\ldots,5$ made up of  $I$ distinct symbols. These patterns are repeated 20 times. For each pattern length $n_l$ and each partition ${\cal A}=\{{\cal A}_1 ,{\cal A}_2,\ldots ,{\cal A}_{I}\}$ of $(1,2, \ldots, n_l)$, 
 one sequence is generated in a way so that elements of each partition subset  ${\cal A}_i$ are symbol $i$'s positions  in the sequence. E.g. for $n_l=5$ and partition ${\cal A}=\{{\cal A}_1 ,{\cal A}_2\}=\{\{1, 3, 5\},\{2,4\} \}$, symbol '$1$' occurs at positions ${\cal A}_1=\{1, 3, 5\}$ and symbol '$2$' occurs at positions ${\cal A}_2=\{2, 4\}$, yielding the symbol sequence ('$1$','$2$','$1$','$2$','$1$').
 \item Audio recordings:
\begin{description}
 \item[Voice:] Informal low quality and short voice recordings of very simplified beat boxing, each consisting of 2-3 different sound categories with different degrees of tonality with a simple changing rhythm, a sequence of a repetitive three-sound pattern, and a ritardando, altogether 5 recordings each of 10-13 s duration. In order to demonstrate the unsupervised character of our system we choose sounds that do not belong to a predefined category (e.g. an acoustical instrument).
  \item[ENST Drums:] Formal high quality and automatically annotated recorded drum sequences. 5 segments described in terms of style, complexity and tempo as disco (simple slow, complex medium), rock (simple fast), country (simple slow, complex medium) \citep{gillet2004}.
 \end{description}
 The audio recordings are annotated, so they can be evaluated. Audio data is available on the website \citep{marxer14web}. 
\end{itemize}
 
\subsection{Results}
\label{sec:eval}

The system architecture consists of the processing chain: 1) {\it onset detection and feature extraction} 2) {\it clustering}, 3)  {\it expectation}.
We will evaluate stages 1), 2), 3) in isolation, 
1) + 2) together (referred to as {\it transcription}) and the entire chain 1) + 2) + 3) together (referred to as {\it prediction}). We use the repetitive symbol sequences, in order
to assess expectation, i.e.\ learning rate and noise robustness of the sequence analysis. The audio recordings are used to test the processing stages of the entire system separately. 
\begin{figure}
 \centering
 \includegraphics[keepaspectratio=true,width=0.49\textwidth]{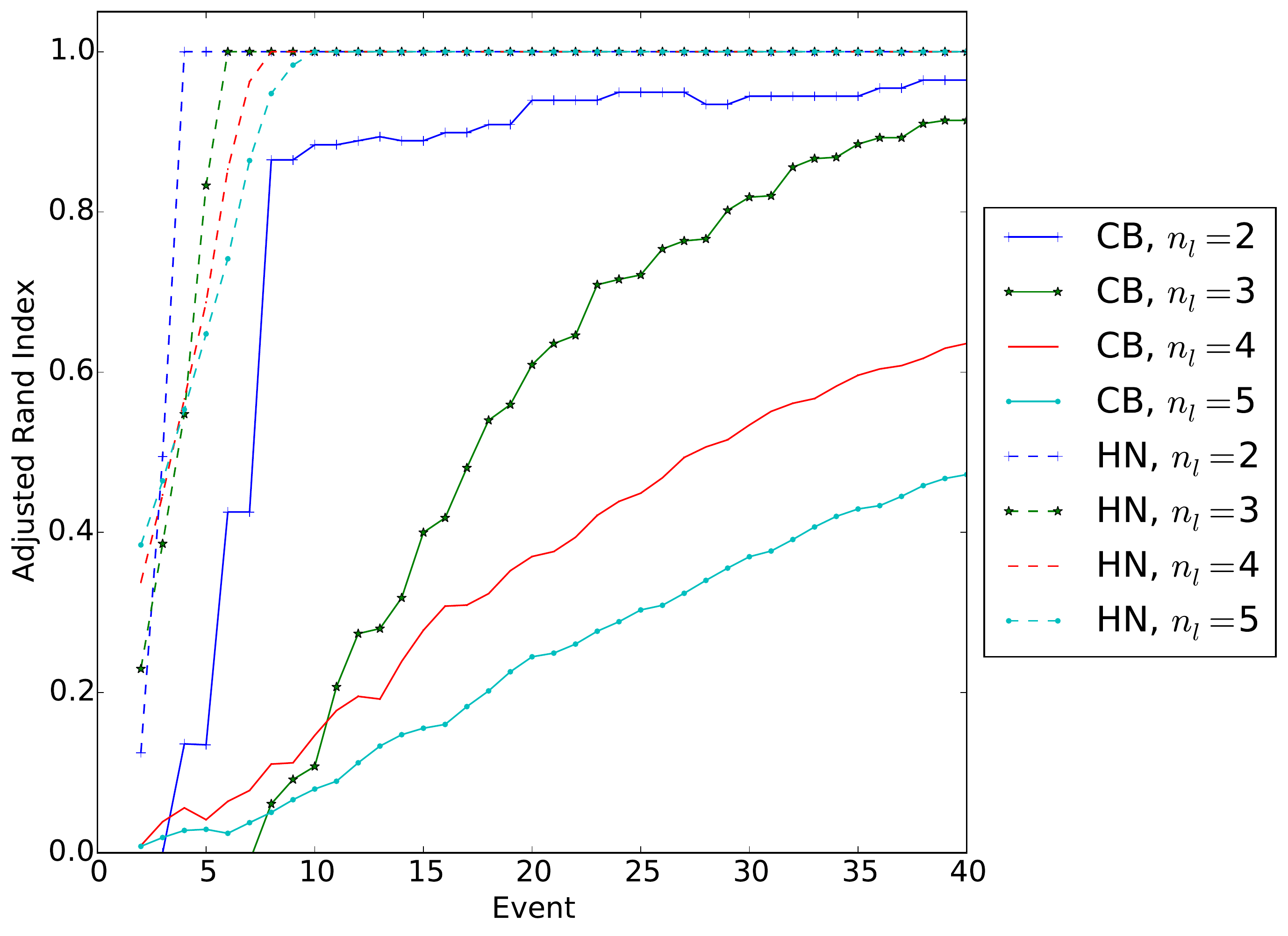}
 \caption{Learning rate of the two  sequence learning algorithms (CB and  HN),  depending on the number of pattern repetitions. The ARI  (Eq.~\ref{eqn:ari}) is given for  an increasing number of repetitions of a pattern with various lengths $n_l=2,3,4,5,6.$ HN reaches a perfect ARI quickly, in contrast to CB. 
 \comment{squeeze all three pics vertically so they dont take so much space?}
 }
 \label{fig:patternLengthRepetitions}
\end{figure}
\subsubsection{Learning Rate and Noise Robustness with Repetitive Symbol Sequences}
We assess the learning rate and noise robustness of the sequence analysis stage (CB and HN). 
The sequence learning algorithm  receives an initial chunk of a repetitive symbol sequence (Section~\ref{sec:data_sets})  as input. From this, the algorithm determines the most probable (expected) next $4 n_l$ symbols.
 The algorithm outputs the expected next symbols 
 $c_{t+1},\ldots, c_{n_l \cdot 4}$,
 given the annotated symbols $a_1, a_2, \ldots, a_{t}$. 
For each $t,$ from the predictions 
 $c_{t+1},c_{t+2},\ldots, c_{n_l \cdot 4}$ 
 a partition ${\cal C}$ is generated and compared with the partition of the corresponding ${\cal A}$ based on annotations, as explained in Section~\ref{sec:clustereval}. For $t\le n_l \cdot 5$ all annotations so far are used
 for prediction, then only the last 12 annotations $a_{t-11},a_{t-10}, \ldots a_t$ are used for prediction. For the stochastic BM, ARI is averaged
 over 100 runs of all partitions of a given length $n_l$. 
 The trivial sequence that consists of a constant repetition of the same symbol is not considered.
First we 
assess how the learning rate scales with \emph{pattern length} and number of \emph{pattern repetitions}.
Fig.~\ref{fig:patternLengthRepetitions} shows the averaged ARI across  
of all partitions of lengths $n_l=2,\ldots 5$.
\comment{how about the beginning lengths?}
 For this test, the HN is set to a maximum $N$-gram length of $N=5$. 
The HN reaches perfect prediction (ARI=1) after $4 n_l$ events (2 pattern repetitions).
CB seems to converge much more slowly  than HN, for $n_l=2$ reaching an ARI of higher than $0.8$ \comment{exact value?} after 8 events, then increasing much more slowly.
 For higher $n_l$, CB seems to converge towards perfect prediction even more slowly.  
 \comment{Does the graph start with 2 repetitions?}

 \begin{figure}
 \centering
 \includegraphics[keepaspectratio=true, width=0.49\textwidth]{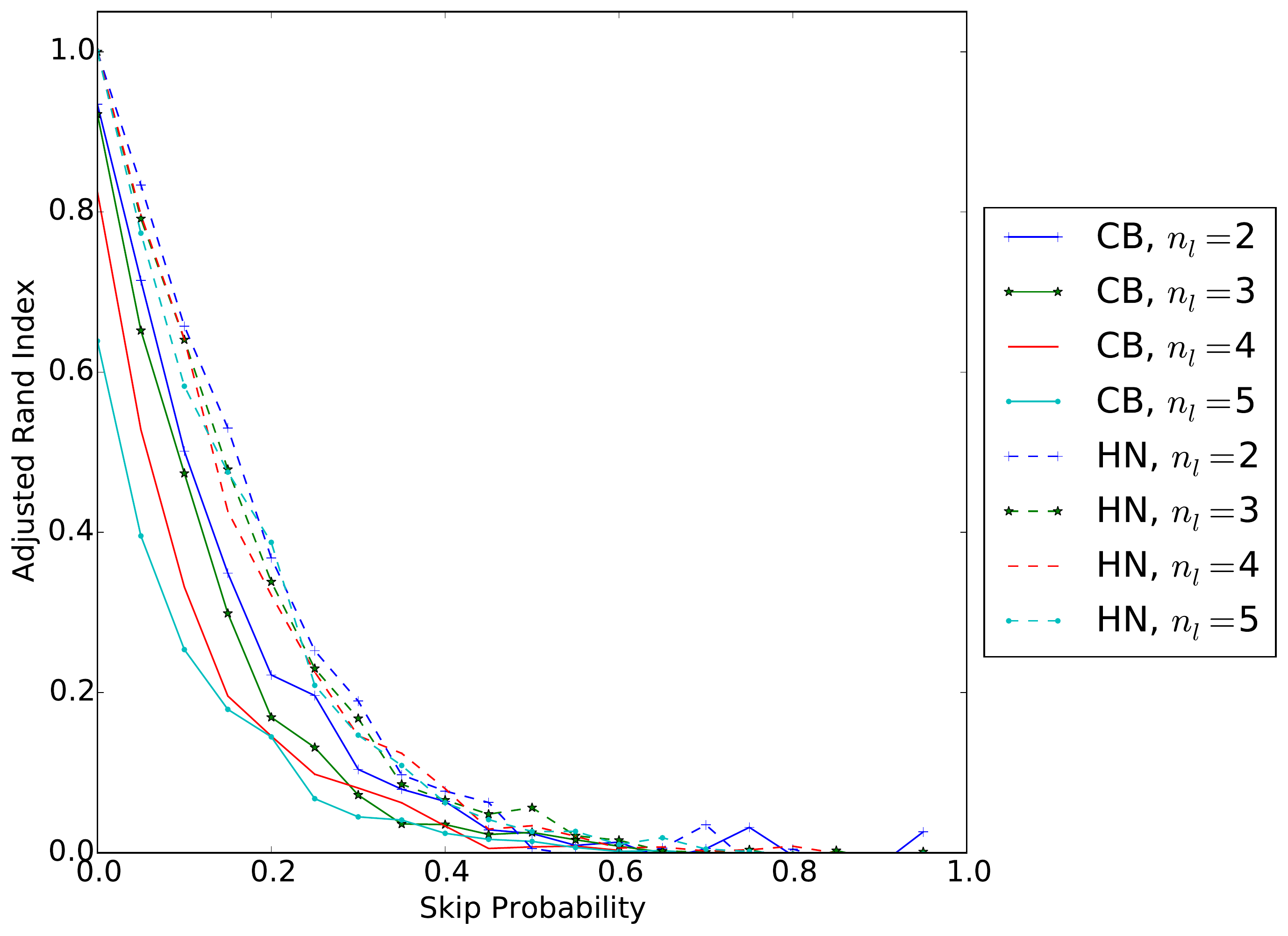}
 \caption{Robustness of the two sequence learning algorithms (CB, HN, cf. Fig. \ref{fig:patternLengthRepetitions}) with respect to skipping noise. The ARI  is given for  a sequence of 20 repetitions of  patterns of different lengths $n_l$ and increasing probability $p_{sk}$ of randomly skipping an event. For $p_{sk} < 0.4,$ HN performs better than CB.  
}
 \label{fig:patternLengthSkip}
\end{figure}  
Different types of noise are used to transform the sequence in order to assess the robustness of the sequence learning techniques:
\begin{description}
 \item[Skipping noise:] 
 In the original sequence, a symbol is skipped  with a given probability $0\le p_{sk} \le 0.95$. 
 \item[Switching noise:] 
 In the original sequence, with a given probability of $0\le p_{sw} \le 0.95$, a symbol is selected randomly with uniform distribution across the $n_l$ alternative symbols. 
\end{description}

The average ARI is calculated  over 100 runs for $n_l=2,3$, over 50 runs for $n_l=4$ and 20 runs for $n_l=5$ for both CB and HN. 
Fig.~\ref{fig:patternLengthSkip} shows how the prediction performance (ARI) is affected by skipping symbols  with a defined $p_{sk}$ in the repetitive symbol sequences. This simulates e.g.\ the failure of  the onset extraction algorithm to detect an event.  The prediction is performed, given a sequence of 20 repetitions of the basic pattern. For HN and CB, the performance degrades until $p_{sk}=0.5$, where random guess level is reached (ARI=$0$). 
Until $p_{sk}=0.4$, HN appears to be more robust towards skipping noise than CB, with CB having a worse ARI for higher $n_l$. 

\begin{figure}
 \centering
 \includegraphics[keepaspectratio=true, width=0.49\textwidth]{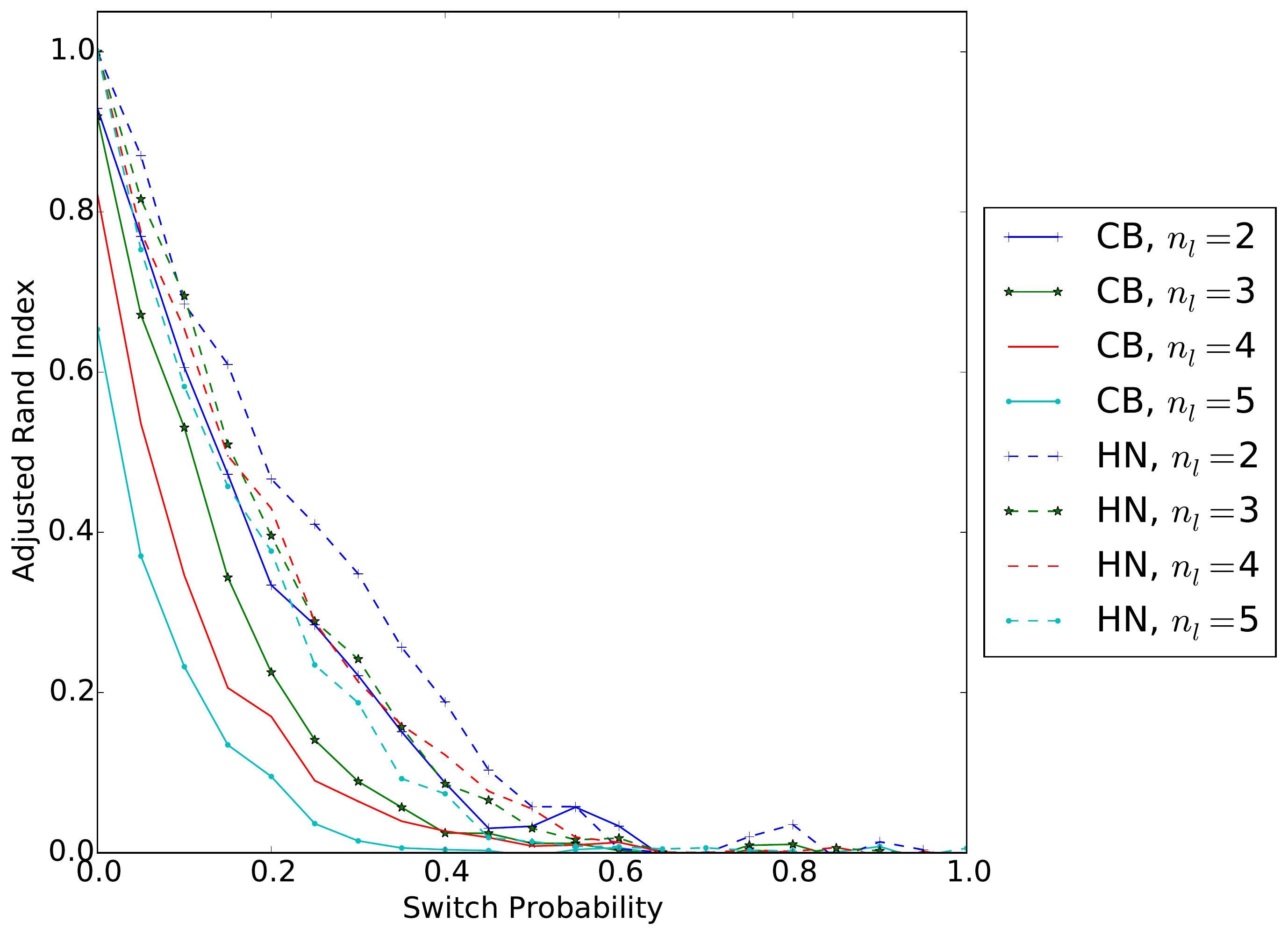}
 \caption{Robustness of CB and HN (cf.\ Fig. \ref{fig:patternLengthRepetitions}) with respect to switching noise. The ARI  is given for  an 
 increasing probability $p_{sw}$ of randomly switching a symbol. HN performs better than CB for $p_{sw}<0$, reaching random guess level (ARI=0) for $p_{sw}=0.5$.}
 \label{fig:patternLengthSwitch}
\end{figure}

In Fig.~\ref{fig:patternLengthSwitch}, the effect  of clustering errors on the sequence learning process is simulated.  With increasing switching probability $p_{sw},$ a symbol is replaced by any of the $n_l$ symbols under uniform distribution. The graph shows the prediction performance using the ARI for CB and HN  for different pattern lengths. 
The results are similar as for skipping noise (Fig.~\ref{fig:patternLengthSkip}): HN is more robust wrt noise
than BM, reaching random guess level (ARI=$0$) for $p_{sw}=0.5.$ 
It can be summarized that for relatively small noise the HN appears to be more robust to skipping and switching noise, especially for  longer pattern lengths.


\subsubsection{Testing of Processing Stages with Audio Recordings}
The tests with the \emph{Voice} recordings (Section~\ref{sec:data_sets}) serve as a proof of concept of clustering with dynamically varying numbers  of clusters.
The \emph{ENST} recordings are used for a more comprehensive quantitative evaluation of the system.
We test each process stage separately. For the audio, the sample rate is $f_s=44100$ Hz. For segmentation and feature extraction, the hop size is $128$ samples, and the window size is $1024$ samples.

\paragraph{Onset detection}
For the evaluation of the onset detection (Section~\ref{sec:preprocessing}) we employ a widely used procedure \citep{downie05,leveau04}. Onset times manually annotated by subjects serve as references. The onsets estimated by the onset detection
algorithm are then compared to the manually annotated onsets.  Annotated and estimated onsets are considered a match when their difference in time is smaller than a given threshold.  In our evaluation, we use an onset match threshold of $\sim50$\ ms. Since the data is assumed to be monophonic, the evaluation only permits a one-to-one mapping between estimated and annotated onsets.

Using the following onset detection parameters: smoothing length $M=33$ in Eq.~\ref{eqn:odfsmoothing},  sensitivity $C=0.9$ 
 and look-ahead window length $P=10$ 
 in Eq.~\ref{eqn:odfthreshold},   threshold window length $ W=11$ in Eq.~\ref{eqn:max_window},  and silence threshold $\theta_s=0.002$ in Eq.~\ref{eqn:silence_threshold},  onset detection yields an F-measure of $\sim 99$\% for  the {\it Voice} data set. 
Therefore, we focus 
on the clustering and the prediction stage.  We also notice that for smoothing lengths \comment{is that the $M$?} $M>33$ the system does not improve significantly.  Large smoothing lengths reduce the temporal precision of onsets, which is important for good feature extraction, since most of the information about an event is located in the attack.


\paragraph{Clustering} \label{par:clustering}
We now compare the performance of our incremental online Cobweb clustering and benchmark offline (batch) HDP-HMM clustering with a constant yet inferred cluster number as a benchmark. 
In order to assess the clustering process in isolation, we assume error-free
onset detection on the previous stage.  In order to achieve this, we use the annotated onsets as input. In order to assess the stability of the system, we tested it performing a grid search on 
the two most sensible parameters involved in the task and the algorithm. For Cobweb, we explore the analysis window length $L$ (Section~\ref{sec:timbre_rep}) and the acuity $a$ (Eq.~\ref{eqn:cobweb}). 
On the parameter grid, the window length/acuty pair with maximal ARI is determined, extending the parameter grid if the maximum lies on the grid border, with empirically set constant grid step sizes. 

For \emph{Voice}, Cobweb performance peaks at  ARI=$82.7\%$ for $L=150 ms,a=18.5$ on a parameter grid over 
$L=50, 75, \ldots, 175;  a=15, 15.5, \ldots, 19$. 
For \emph{ENST}, Cobweb performs best  at  $85.7\%$ for $L=50 ms,a=13.5$ on a parameter grid over 
$L=25, 50, \ldots, 100;  a=13, 13.5, \ldots, 15$. (cf.\ Tables~\ref{tab:clustFrecall} and \ref{tab:clustFmeasureENST} in the supplementary material\cite{marxer14web})
This means that the timbre model and clustering process can successfully classify the audio events.  We also notice that \emph{Voice} needs a much longer analysis window than \emph{ENST}.
\comment{How many Mfcc features, which Mfcc implementation (since there are so many differences in implementations); update explanation for result}
This test, as explained above, was performed using the annotated onsets.  The results could change when the onsets are estimated.  This effect is evaluated in the transcription test (Section ~\ref{sec:transcription}).

For the HDP-HMM, we first reduce  the feature vectors of the input  to $D$ dimensions by means of a PCA on the full sequence. The observation distributions used are Gaussian with parameters sampled i.i.d. from a normal inverse Wishart prior  \cite{johnson2013hdphsmm}
with parameters $\mu_0=\mathbf{0}, \kappa_0=0.4,\Lambda_0=0.001,\nu_0=D+2$.\footnote{Cf.~\citet{murphy_conjugate_2007}, Section~9.2., p.~20 for the meaning of the parameters.}
\comment{variables need to be explained} The maximum number of states of the weak limit approximation inference is set to $10$ and the number of Gibbs sampling iterations to $100$.
For \emph{Voice}, HDP-HMM performance peaks at  ARI=$99.1\%$ for $\gamma=8.0,\alpha=7.0,D=2$ on a parameter grid over $\gamma,\alpha=4.0, 5.0, \ldots, 11.0$ and $D={2,3}$.
For \emph{ENST}, HDP-HMM performs best at  ARI=$84.0\%$ for HDP concentration parameters $\gamma=6.0,\alpha=12.0$ \cite{teh_hierarchical_2010} 
and $D=2$ on a parameter grid over 
$\gamma,\alpha=4.0, 5.0, \ldots, 13.0$ and $D={2,3}$. 
Benchmark HDP-HMM performs  better for \emph{Voice} than Cobweb, whereas
for \emph{ENST}, Cobweb performs $1.7\%$  better than HDP-HMM. When comparing these results one has to keep in mind that HDP-HMM clustering
has learned offline jointly a stable cluster number and the transition probabilities, exploiting sequential information whereas Cobweb has been trained online with an adaptive cluster number. 

\paragraph{Transcription}\label{sec:transcription}
The transcription test evaluates the subsystem composed of onset detection, feature extraction, and clustering.  In contrast to the expectation test, the entire symbol (inter-onset interval) sequence   $c_1,c_2, \ldots, c_t$ extracted from the clustering stage is always used from the beginning to predict the next symbol (inter-onset interval) $c_{t+1}.$
The annotations $\mathbf{a}$ are not used for prediction, only for evaluation. The partitions generated from the detected events were compared with the partitions generated from the annotated labels using ARI. Online learning Cobweb with dynamically changing clustering numbers and offline learning HDP-HMM with a constant cluster number are compared. For Cobweb,  {\it Voice} performs with $ARI=81.3 \%$ for $L=150,a=17$.   On the {\it ENST} data set Cobweb yields $ARI=76.3\%$ for $L=50,a=13.5$, using the same parameter grids as for the clustering (p.~\ref{par:clustering}). In comparison to the results for clustering, the $ARI$ degrades a bit in particular for \emph{ENST} due to wrongly estimated onsets. 
(cf.\ Tables ~\ref{tab:transVoice} and \ref{tab:transENST} in the supplementary material \cite{marxer14web})
HDP-HMM transcription performance for \emph{Voice} peaks at  ARI=$98.8\%$ for $\gamma=8.0,\alpha=12.0$ on a parameter grid over 
	$\gamma,\alpha=4.0, 5.0, \ldots, 13.0$. 
	For \emph{ENST}, HDP-HMM transcription performance peaks at  ARI=$76.2\%$ for $\gamma=5.0,\alpha=8.0$ on the same parameter grid  as for \emph{Voice}. 
For \emph{ENST}, HDP-HMM and Coweb are almost equal. 
	Although for \emph{Voice}, the ARI is much higher for the HDP-HMM benchmark than for Cobweb, we have to keep in mind that  Cobweb learns online with changing cluster numbers over time whereas HDP-HMM is trained offline with a constant number of clusters.

\paragraph{Expectation}
The expectation test evaluates the performance of the sequence learning module on the data sets.  We predict the cluster label $c_{t+1}$ of event $t+1$ based on the annotations from the start:
$a_1,a_2,\ldots, a_{t}.$ 
\begin{table}[tb]
\centering
\caption{Expectation of {\it Voice} (left) and {\it ENST} (right): ARI (in \%) for different maximum lengths $N$ of CB/HN (rows). \comment{hp: boltzmann for n=1 best?}}
\begin{tabular}{@{}c|cc@{}}\hline
$N$ & CB & HN \\\hline
2 & 7.4 & 22.4 \\
3 & 6.8 & 27.3 \\
4 & 7.3 & 41.1 \\
5 & 5.1 & { \bf 50.9 } \\
6 & 4.4 & 50.9 \\
7 & 5.1 & 50.9 \\
\hline
\end{tabular}
\quad\quad\quad
\begin{tabular}{@{}c|cc@{}}\hline
$N$ & CB & HN \\\hline
2 & 6.0 & 18.9 \\
3 & 9.1 & 28.9 \\
4 & 7.8 & { \bf 43.2 } \\
5 & 6.3 & 42.7 \\
6 & 6.6 & 42.7 \\
7 & 7.7 & 42.6 \\
\hline
\end{tabular}
\label{tab:expecVoiceENST}
\end{table}
Results in Table~\ref{tab:expecVoiceENST} show that for the prediction of the  sequences of the {\it Voice} and the {\it ENST} data set, HN ($ARI=43.2\%$ for $N=4$) works a lot better than CB, \hl{which yields an $ARI=7.8\%$, just slightly better than random ($0\%$). CB's low performance can be attributed to various factors:  In general, many traditional recurrent networks are known to have a slow learning rate.}\citep{hochreiter_long_1997} \hl{In particular, we have observed slow learning rate} (Fig.~\ref{fig:patternLengthRepetitions}) \hl{and low noise robustness} 
(Fig.~\ref{fig:patternLengthSkip} \& Fig.~\ref{fig:patternLengthSwitch}). \hl{
Whereas for HN, the updates in frequency 
counts are getting smaller relative to the count so far (from 1 to 2 is a higher step relative to 1 than from 100 to 101 relative to 100), in CB the weights are updated with a constant learning rate $\mu$} (Eq. \ref{eqn:contrast_div}). \hl{Weight updates are performed by stochastic gradient descent where each instance is only used once when it has just occurred. Although this is cognitively plausible if we assume that only a limited number of instances can be stored by the cognitive system, it comes with the price of diminished learning speed, compared to a system where the update is performed using a batch of instances. In addition, the architecture of the CB may be suboptimal w.r.t. the hidden nodes. 
Also, in the network, new hidden nodes are only generated one at a time, further limiting learning speed. Furthermore, the parameters $\theta_k$ for creating new hidden nodes are chosen heuristically and may be suboptimal for short sequences like the one presented. 
} 
We can also see that for $n$-gram maximum lengths $n>5$ for {\it Voice} ($n>3$ for {\it ENST}) the result does not improve. 
\hl{For linguistic data, slower convergence and worse performance of CB relative to HN is also observed in}  \citet{pfle02}, pp.~80\&133. In the sequel, we will only use HN.

\paragraph{Prediction}
The prediction task consists in running the full system including HN as the sequence analyzer (Tables~\ref{tab:predVoice} and \ref{tab:predENST}).  
After the transcription of the events $c_1,\ldots, c_{t},$ the system predicts the next symbol and the timing of it (the next IOI) $c_{t+1}.$ For evaluating the match between predicted and annotated onsets, we set the  tolerance threshold to 150~ms. For the best configuration, the full prediction yields an ARI of $27.2\%$ for \emph{Voice} and an ARI of $39.2\%$ for \emph{ENST}. The performance is limited by the weakest performance of its components, in this case the sequence analysis.

\begin{table}[tb]
\centering
\caption{Full Prediction for the {\it Voice} data set using HN: ARI (in \%) for different temporal acuities $a_t$ from Eq.~\ref{eqn:cobweb} (rows) and timbral acuities $a$ from Eq.~\ref{eqn:cobweb} (columns).  }
\begin{tabular}{@{}c|ccccccc@{}}\hline
$a_t \backslash a$
          &        17   &      17.5   &        18   &      18.5   &        19   &      19.5   &        20   \\\hline
     0.05 &      16.7   &      15.5   &      16.3   &      16.0   &      25.2   &      23.8   &      23.8   \\
   0.0625 &      15.9   &      15.3   &      16.2   &      16.0   &      25.2   &      24.0   &      24.0   \\
    0.075 &      14.4   &      14.3   &      15.2   &      16.7   & {\bf 27.2 } &      25.8   &      25.8   \\
   0.0875 &      15.7   &      16.4   &      17.2   &      17.3   &      25.8   &      24.5   &      24.5   \\
\hline
\end{tabular}
\label{tab:predVoice}
\end{table}

\begin{table}[tb]
\centering
\caption{Full Prediction for the {\it ENST} data set using HN: ARI (in \%) for different temporal acuities $a_t$ from Eq.~\ref{eqn:cobweb} (rows) and timbral acuities $a$ from Eq.~\ref{eqn:cobweb} (columns).}
\begin{tabular}{@{}c|ccccccc@{}}\hline
$a_t \backslash a$
          &        18   &      18.5   &        19   &      19.5   &        20   &      20.5   &        21   \\\hline
    0.075 &      33.1   &      33.8   &      33.0   &      32.6   &      36.2   &      34.7   &      33.3   \\
   0.0875 &      35.1   &      36.2   &      35.3   &      34.2   &      37.8   &      36.3   &      34.9   \\
      0.1 &      36.3   &      37.6   &      36.6   &      35.4   & {\bf 39.2 } &      37.6   &      36.2   \\
   0.1125 &      35.9   &      37.2   &      36.2   &      35.0   &      38.8   &      37.2   &      35.8   \\
    0.125 &      34.6   &      35.9   &      34.9   &      33.7   &      37.5   &      35.9   &      34.5   \\
\hline
\end{tabular}
\label{tab:predENST}
\end{table}

\subsection{Examples}
In this section, we present a few examples (audio on the website \citep{marxer14web}) of transcription and prediction using HN in order to demonstrate the performance, evolution and shortcomings of the system. From \citet{hazan2009}, we adopt the procedure to optimally map the annotated symbols to the clusters found by the clustering algorithm. We calculate the matching matrix between the annotations ('score') and clusters of each event. In this matching matrix, we then iteratively yield the maximal entry, thereby establishing a connection between a row (annotations)
 and a column (clusters). After eliminating the row and column of the maximal entry, we determine the maximal entry again until the matrix vanishes. 
\begin{figure}[htb]
 \centering
 \includegraphics[keepaspectratio=true, width=0.49\textwidth]{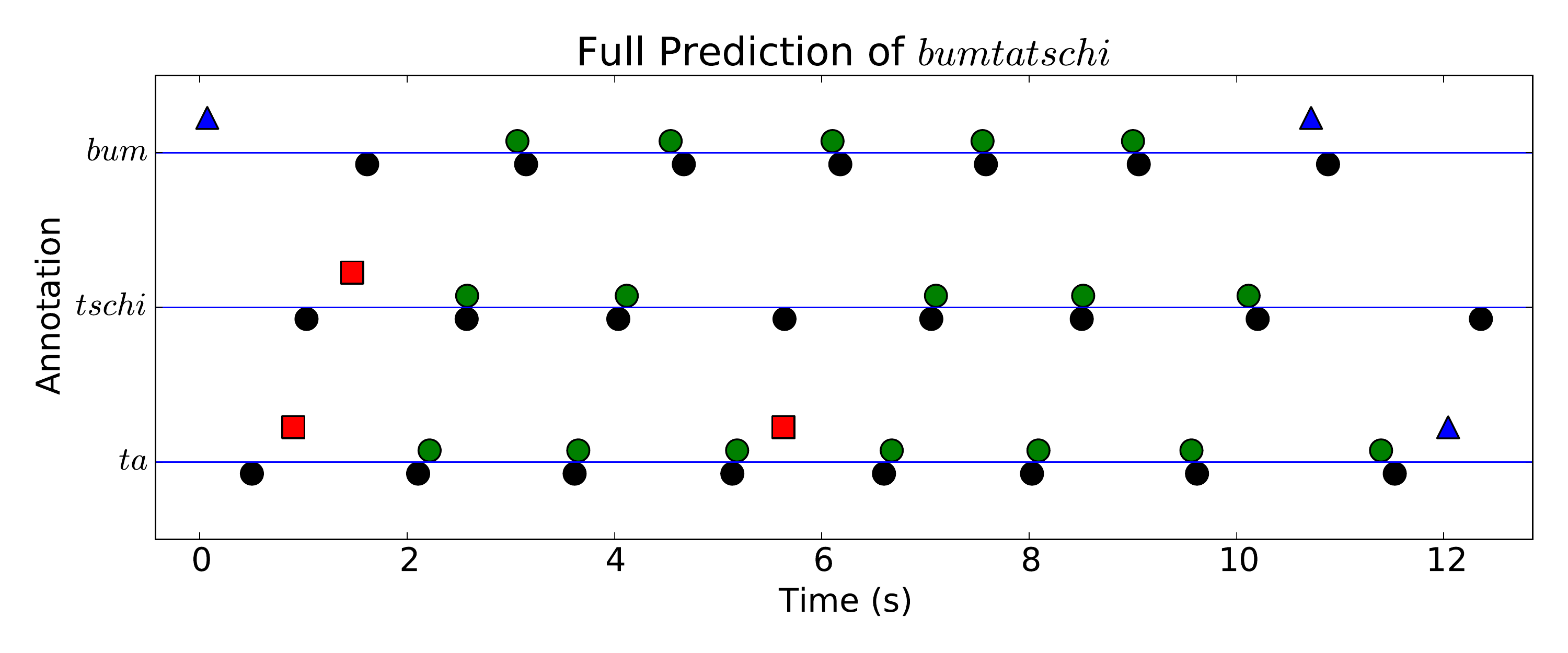}
 \caption{The system (with HN) quickly captures a simple {\it ta-tschi-bum} pattern. Time (horizontal axis) is mapped versus event labels (one line each for \emph{'ta', 'tschi',} and \emph{'bum'}). Annotated labels are indicated in black below the lines. Above the horizontal lines we find events that are correctly estimated ('{\color{green}$\bullet$}'),  matched to the wrong cluster ('{\color{red}$\blacksquare$}'), and unmatched ('{\color{blue}$\blacktriangle$}') due to a wrongly estimated onset.}
 \label{fig:exampleBumtatschi}
\end{figure}
\begin{figure}[tb]
 \centering
 \includegraphics[keepaspectratio=true, width=0.49\textwidth]{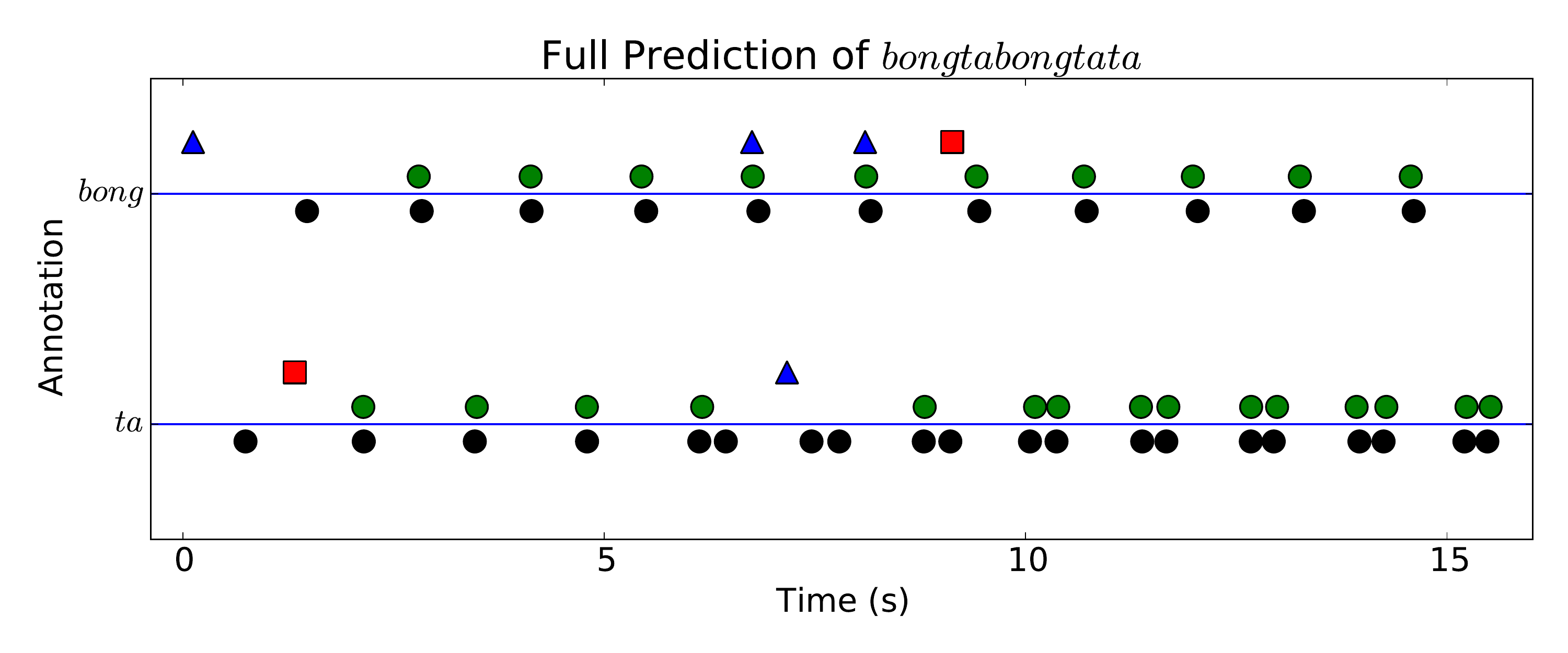}
 \caption{The system (with HN) adapts to a pattern change from {\it ta-bong} to {\it ta-ta-bong} (cf.\ Fig.~\ref{fig:exampleBumtatschi}). }
 \label{fig:exampleBomtschibomtschitschi}
\end{figure} 

 In Fig.~\ref{fig:exampleBumtatschi} and \ref{fig:exampleBomtschibomtschitschi}, we display sequences of annotations and clusters on the same line if they are linked through this mapping. 
In Fig.~\ref{fig:exampleBumtatschi}, a simple {\it ta-tschi-bum} pattern is quickly captured. We can see how the 
first three events annotated as \emph{ta}, \emph{tschi}, and \emph{bum} are 
matched with the wrong clusters \emph{bum} (initial blue triangle above top line), \emph{ta}, and \emph{tschi} (red squares). The first three cluster mismatches are expected, since the system has no previous knowledge of the symbol space nor of the sequence and therefore cannot predict symbols nor patterns that have not yet occurred. At around $5.5 s$, an event annotated as \emph{tschi} is matched
with the \emph{ta} cluster. The timing of the last \emph{bum} is misestimated and for the last
\emph{tschi} timing and cluster matching are wrong.
The time deviation errors are due to the fact that the recorded voice does not follow a temporally regular pattern. 

In Fig.~\ref{fig:exampleBomtschibomtschitschi}, we observe how the system adapts to pattern changes within the sequence.  For the first two events, the cluster matching is wrong. Then, after having processed enough sounds, the system performs correct predictions. In the middle, around $7.5 s$, when the repetition pattern  of \emph{ta} is introduced, for three \emph{ta} events the onset is wrongly estimated, two of these events as well being mismatched with the wrong cluster, and one additional event being only mismatched with the wrong cluster. The errors in the middle of the sequence are due to the pattern change.  The $N$-gram is able to update the statistics and perform correct predictions after three occurrences of the new pattern.  

\begin{figure}[t]
 \centering
 \includegraphics[keepaspectratio=true, width=0.45\textwidth]{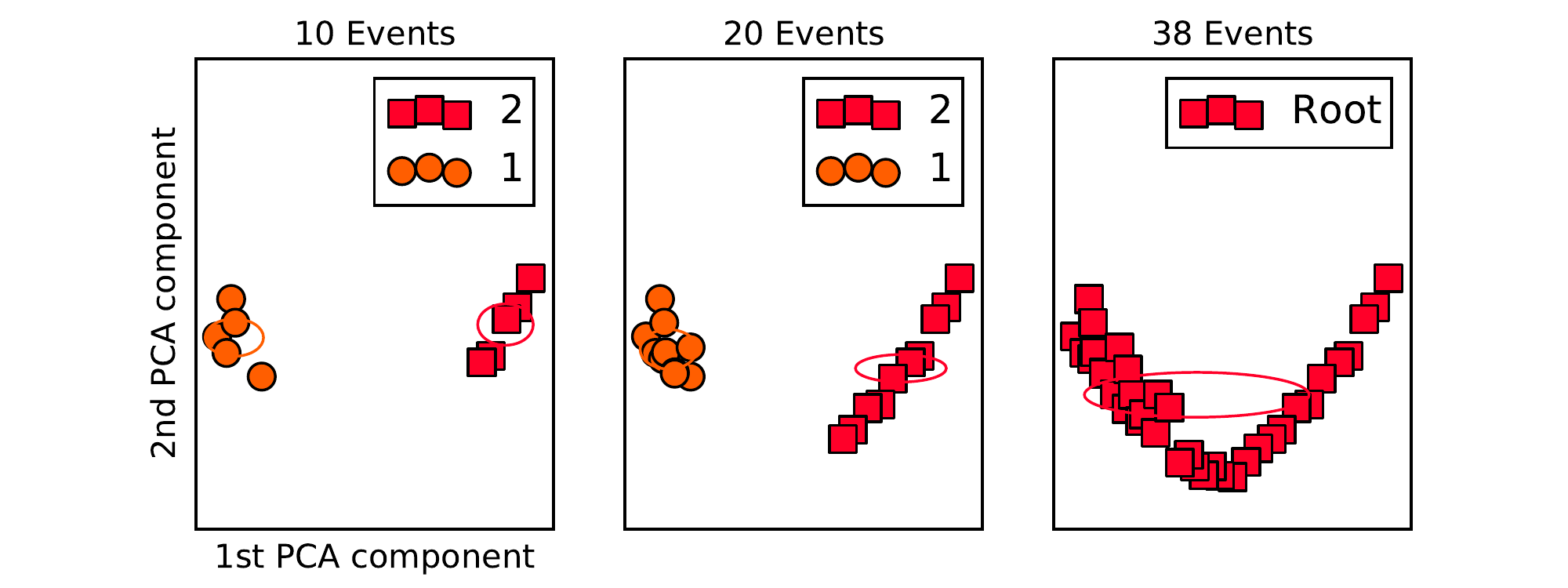}
 \includegraphics[keepaspectratio=true, width=0.45\textwidth]{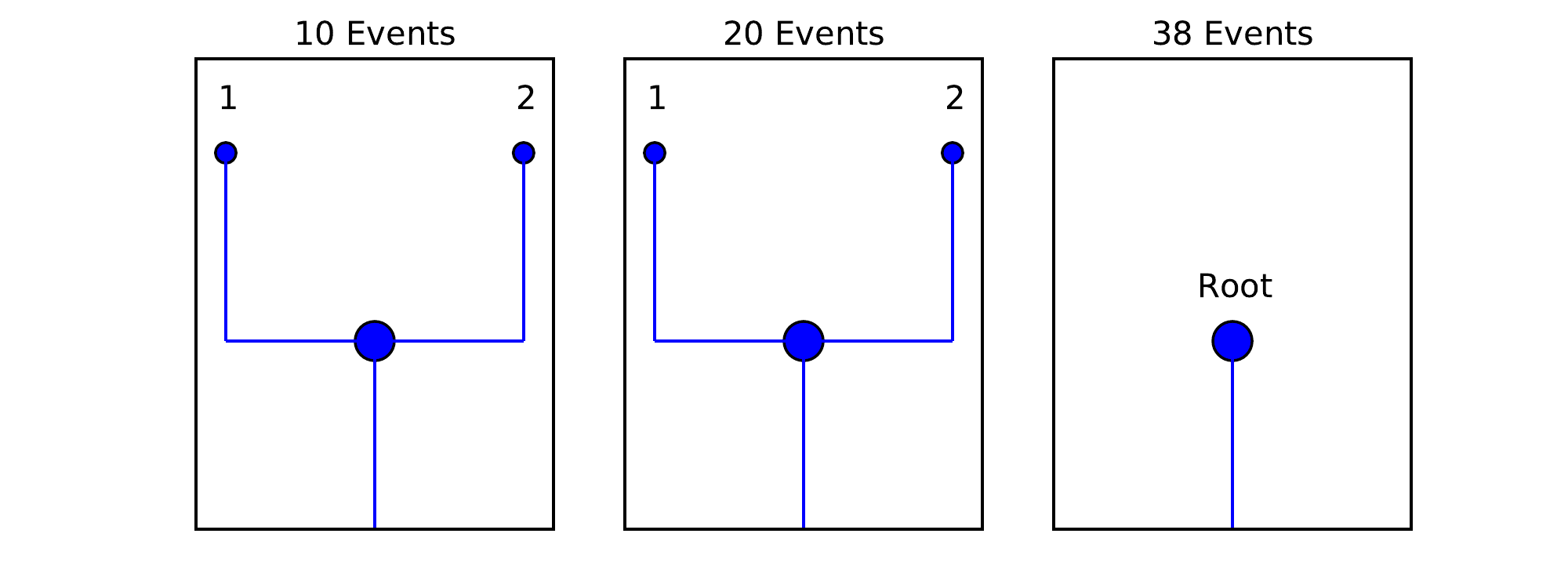}
 \caption{Cluster merging: After 38 events, two clusters ('{\color{orange}$\bullet$}', '{\color{red}$\blacksquare$}') merge into one cluster ('{\color{red}$\blacksquare$}'). The projection of the MFCC vectors (timbre representation) onto their first two principal components (\emph{above}) and the incremental clustering tree (\emph{below}) are shown.}
 \label{fig:mergeHistory}
\end{figure}  
In Fig.~\ref{fig:mergeHistory} (sound and video on the website \citep{marxer14web}),   a sequence of alternating bass drum and hi-hat samples is played. During the sequence, the hi-hat is gradually mixed in a linear fashion with an increasing amount of bass drum and vice versa so that in the end both hi-hat and bass drum are mixed together in a balanced way, yielding a repetitive sequence of similar sounds.
The system recognizes the two sound clusters in the beginning, and finally merges the two clusters into one single cluster. 

\begin{figure}[t]
 \centering
 \includegraphics[keepaspectratio=true, width=0.45\textwidth]{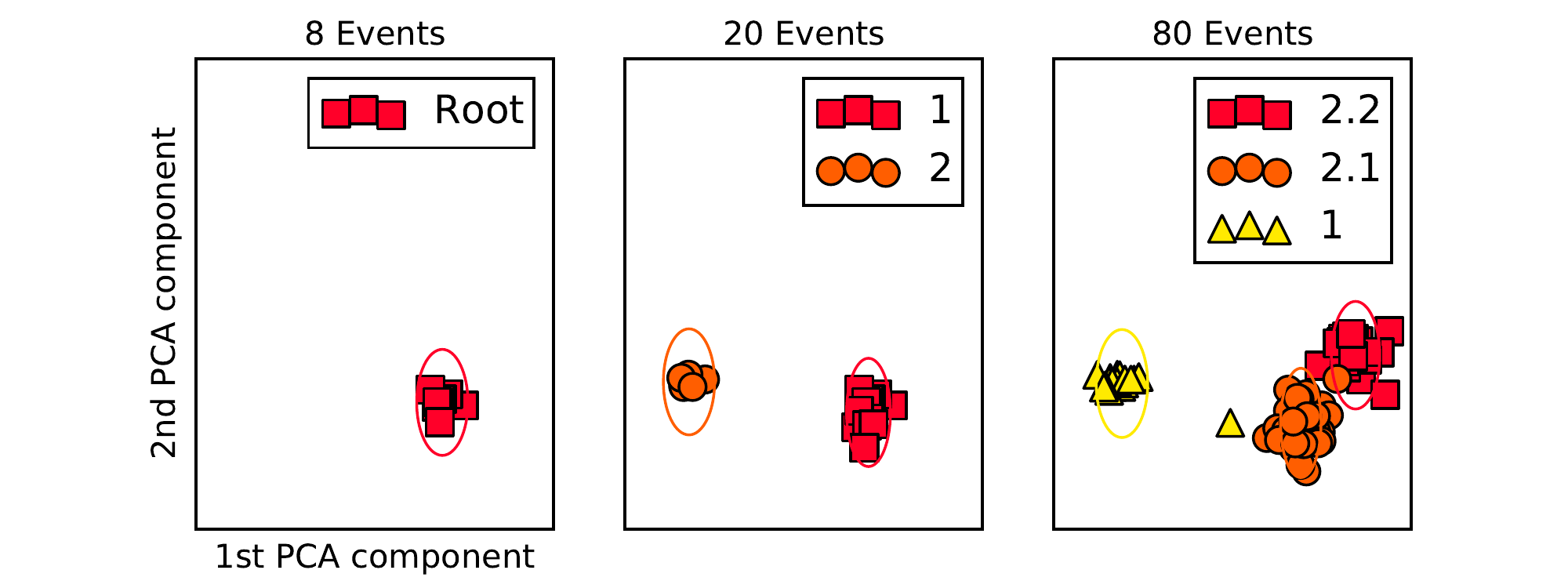}
 \includegraphics[keepaspectratio=true, width=0.45\textwidth]{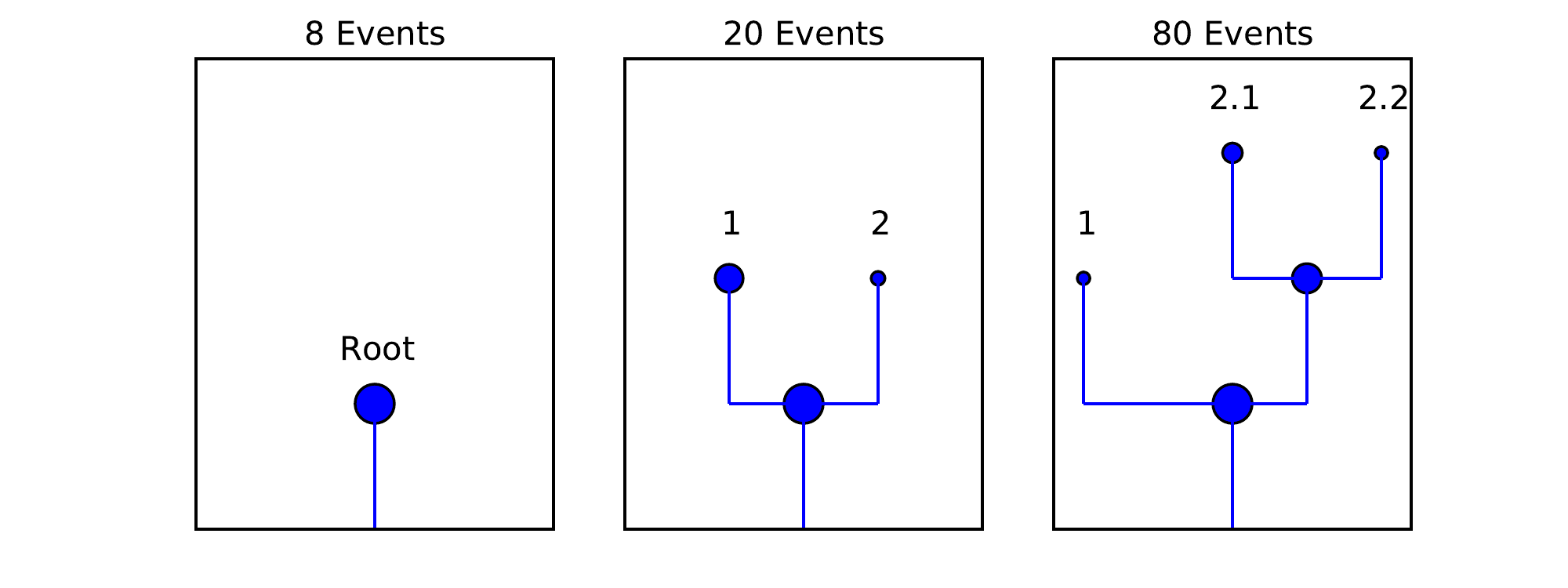}
 \caption{Creation of new clusters: After 20 and 80 sound events,  new clusters ('{\color{orange}$\bullet$}', {'\color{yellow}$\blacktriangle$}') emerge on the fly. Cf.\ Fig. \ref{fig:mergeHistory}.}
 \label{fig:splitHistory}
\end{figure}

In Fig.~\ref{fig:splitHistory} (sound and video on the website \citep{marxer14web}), a sequence of sound events is analyzed that starts with one sound, later
joined by a second and third sound. The system is able to split the initial cluster gradually into 2 and 3 clusters.

\section{Conclusion and Perspectives}
We have presented a full system that predicts the next sound event from the previous events, operating on audio data.  Taking into account no previous knowledge,  neither on the used sounds or instruments nor on the timing and rhythmical structure of the audio segment, the system starts from {\it tabula rasa}, performing predictions from the very first sound event.  The system  adapts to pattern changes in the sequence as well as the appearance of new sounds or instruments at any time.
Currently the system is limited by the lack of metrical analysis, making it
especially sensitive to  missed onsets. 
Considering the metrical context could significantly improve the quality of  predictions.  For this goal, a metrical alignment procedure\citep{marchini2010a,marchini2011unsupervised} could be combined with incremental learning.
As alternatives to CB and HN,  variable length Markov models \citep{Buhlmann99variablelength,marchini2010a,marchini2011unsupervised} \hl{or other deep learning architectures can be used, thereby overcoming the context length limitation of HN and CB and the slow learning of CB. 
The  long short-term memory (LSTM)} \citep{hochreiter_long_1997} \hl{is a recurrent neural network that had been developed to capture dependencies between disconnected distant chunks within the same time series. Crucial to this and for speeding up learning, in LSTM, special memory cells are used. The access to the latter can be opened and closed  by special gating units. Successfully applied to protein homology detection, automatic composition} \citep{eck2002finding}, \hl{handwriting and spoken language recognition, LSTM could be used to replace CB or HN and improve learning speed in our application.}   
HDP-HMM \cite{teh_hierarchical_2010} could be adapted to online learning \cite{bargi2012} with incremental addition/removal of clusters comprising also segmentation, thereby replacing onset detection.
\comment{It can be useful to integrate an earlier represenation of the sound into
the BC (not just starting from the discretization) and also include temporal evolution. This way to have a more integrated holistic architecture. It could be better to 
introduce more hidden layers into the BC or use an algorithm such as the Long-Short-Term memory.}
The presented system can also be modified to learn melodies and chord progressions. For learning melodies, in the feature extraction stage (Section~\ref{sec:timbre_rep}), MFCCs need to be replaced by a  pitch detection method as used  in  \citet{marx07} for learning songs by the Mbenzele pygmies or as in \citet{cherla_automatic_2013} for learning guitar riffs. When analysing (piano) chord progressions, MFCCs can be replaced by constant Q profiles \cite{purwins_new_2000,kosta2012unsupervised}. Future work includes the development of a better representation of pitch and harmony, using a larger training set when processing more complex music.

Inspired by these ideas, we imagine a musical improvisational dialogue between a human and a machine in which the human may spontaneously articulate novel ideas such as new sounds, motifs, rhythms, or harmonies. A dumb and ignorant machine would dampen and finally stop the musical flow. But if the machine could take up the novel idea, reply to it, varying the suggestions of its human partner, they could develop an enhanced musical conversation. 

\bibliographystyle{plainnat}
\bibliography{IEEEabrv,emcap07}

\begin{thebibliography}{48}
\providecommand{\natexlab}[1]{#1}
\providecommand{\url}[1]{\texttt{#1}}
\expandafter\ifx\csname urlstyle\endcsname\relax
  \providecommand{\doi}[1]{doi: #1}\else
  \providecommand{\doi}{doi: \begingroup \urlstyle{rm}\Url}\fi

\bibitem[Ackley et~al.(1985)Ackley, Hinton, and Sejnowski]{ackley1985learning}
David~H Ackley, Geoffrey~E Hinton, and Terrence~J Sejnowski.
\newblock A learning algorithm for {B}oltzmann machines.
\newblock \emph{Cognitive science}, 9\penalty0 (1):\penalty0 147--169, 1985.

\bibitem[Assayag and Dubnov(2004)]{assayag2004}
G\'erard Assayag and Shlomo Dubnov.
\newblock Using factor oracles for machine improvisation.
\newblock \emph{Soft Computing}, 8\penalty0 (9):\penalty0 604--610, 2004.

\bibitem[Bargi et~al.(2012)Bargi, Xu, and Piccardi]{bargi2012}
A.~Bargi, R.Y.D. Xu, and M.~Piccardi.
\newblock An online {HDP-HMM} for joint action segmentation and classification
  in motion capture data.
\newblock In \emph{Computer Vision and Pattern Recognition Workshops, IEEE
  Computer Society Conf. on}, pages 1--7, June 2012.

\bibitem[Bello and Sandler(2003)]{bell03}
Juan~Pablo Bello and Mark~B. Sandler.
\newblock Phase-based note onset detection for music signals.
\newblock \emph{Proc. IEEE Int. Conf. Acoustics, Speech and Signal Processing},
  5\penalty0 (2):\penalty0 441--444, 2003.

\bibitem[Bertin-Mahieux et~al.(2010)Bertin-Mahieux, Weiss, and
  Ellis]{bertin2010clustering}
Thierry Bertin-Mahieux, Ron~J Weiss, and Daniel~PW Ellis.
\newblock Clustering beat-chroma patterns in a large music database.
\newblock In \emph{Proc. Int. Society Music Information Retrieval}, pages
  111--116, 2010.

\bibitem[Buhlmann and Wyner(1999)]{Buhlmann99variablelength}
Peter Buhlmann and Abraham~J. Wyner.
\newblock Variable length {M}arkov chains.
\newblock \emph{Annals of Statistics}, 27:\penalty0 480--513, 1999.

\bibitem[Cherla et~al.(2013)Cherla, Purwins, and
  Marchini]{cherla_automatic_2013}
Srikanth Cherla, Hendrik Purwins, and Marco Marchini.
\newblock Automatic phrase continuation from guitar and bass guitar melodies.
\newblock \emph{Computer Music Journal}, 37\penalty0 (3):\penalty0 68--81,
  2013.

\bibitem[Conklin and Witten(1995)]{conklin1995multiple}
Darrell Conklin and Ian~H Witten.
\newblock Multiple viewpoint systems for music prediction.
\newblock \emph{Journal of New Music Research}, 24\penalty0 (1):\penalty0
  51--73, 1995.

\bibitem[Downie et~al.(2005)Downie, West, Ehmann, and Vincent]{downie05}
J.~Stephen Downie, Kris West, Andreas~F. Ehmann, and Emmanuel Vincent.
\newblock The 2005 music information retrieval evaluation exchange ({MIREX}
  2005): Preliminary overview.
\newblock In \emph{Proc. Int. Society Music Information Retrieval}, pages
  320--323, 2005.

\bibitem[Duxbury et~al.(2003)Duxbury, Bello, Davies, and Sandler]{duxb03}
C.~Duxbury, J.~Bello, M.~Davies, and M.~Sandler.
\newblock Complex domain onset detection for musical signals.
\newblock In \emph{Proc. Digital Audio Effects Workshop}, London, UK, 2003.

\bibitem[Eck and Schmidhuber(2002)]{eck2002finding}
Douglas Eck and J{\"u}rgen Schmidhuber.
\newblock Finding temporal structure in music: Blues improvisation with {LSTM}
  recurrent networks.
\newblock In \emph{IEEE Workshop on Neural Networks for Signal Processing},
  pages 747--756. IEEE, 2002.

\bibitem[Fisher(1987)]{fish87}
Douglas~H. Fisher.
\newblock Knowledge acquisition via incremental conceptual clustering.
\newblock \emph{Machine Learning}, 2\penalty0 (2):\penalty0 139--172, 1987.

\bibitem[Fox et~al.(2011)Fox, Sudderth, Jordan, and Willsky]{fox_sticky_2011}
Emily~B Fox, Erik~B Sudderth, Michael~I Jordan, and Alan~S Willsky.
\newblock A sticky {HDP}-{HMM} with application to speaker diarization.
\newblock \emph{The Annals of Applied Statistics}, pages 1020--1056, 2011.

\bibitem[Fred and Jain(2005)]{fred2005combining}
Ana~LN Fred and Anil~K Jain.
\newblock Combining multiple clusterings using evidence accumulation.
\newblock \emph{Pattern Analysis and Machine Intelligence, IEEE Transactions
  on}, 27\penalty0 (6):\penalty0 835--850, 2005.

\bibitem[Gillet and Richard(2004)]{gillet2004}
O.~Gillet and G.~Richard.
\newblock Automatic transcription of drum loops.
\newblock \emph{Proc. IEEE Int. Conf. Acoustics, Speech, and Signal
  Processing}, 4, 2004.

\bibitem[Gluck and Corter(1985)]{gluc85}
M.~Gluck and J.~Corter.
\newblock Information, uncertainty, and the utility of categories.
\newblock \emph{Proc. Annual Conf. of the Cognitive Science Society}, pages
  283--287, 1985.

\bibitem[Hazan et~al.(2009)Hazan, Marxer, Brossier, Purwins, Herrera, and
  Serra]{hazan2009}
A.~Hazan, R.~Marxer, P.~Brossier, H.~Purwins, P.~Herrera, and X.~Serra.
\newblock What/when causal expectation modelling applied to audio signals.
\newblock \emph{Connection Science}, 21:\penalty0 119 -- 143, 2009.

\bibitem[Hochreiter and Schmidhuber(1997)]{hochreiter_long_1997}
Sepp Hochreiter and J{\"u}rgen Schmidhuber.
\newblock Long {Short}-{Term} {Memory}.
\newblock \emph{Neural Computation}, 9\penalty0 (8):\penalty0 1735--1780, 1997.

\bibitem[Johnson and Willsky(2013)]{johnson2013hdphsmm}
Matthew~J. Johnson and Alan~S. Willsky.
\newblock Bayesian nonparametric hidden semi-{M}arkov models.
\newblock \emph{Journal of Machine Learning Research}, 14:\penalty0 673--701,
  February 2013.

\bibitem[Kosta et~al.(2012)Kosta, Marchini, and Purwins]{kosta2012unsupervised}
Katerina Kosta, Marco Marchini, and Hendrik Purwins.
\newblock Unsupervised chord-sequence generation from an audio example.
\newblock In \emph{ISMIR}, pages 481--486, 2012.

\bibitem[Larsen and Aone(1999)]{larsen1999fast}
B.~Larsen and C.~Aone.
\newblock {Fast and effective text mining using linear-time document
  clustering}.
\newblock In \emph{Proc. ACM SIGKDD Int. Conf. on knowledge discovery and data
  mining}, pages 16--22, 1999.

\bibitem[Lartillot et~al.(2001)Lartillot, Dubnov, Assayag, and
  Bejerano]{lartillot2001}
Olivier Lartillot, Shlomo Dubnov, G\'erard Assayag, and Gill Bejerano.
\newblock Automatic modeling of musical style.
\newblock In \emph{Proc. Int. Computer Music Conference}, 2001.

\bibitem[Leveau and Daudet(2004)]{leveau04}
Pierre Leveau and Laurent Daudet.
\newblock Methodology and tools for the evaluation of automatic onset detection
  algorithms in music.
\newblock In \emph{Proc. Int. Society Music Information Retrieval}, pages
  72--75, 2004.

\bibitem[Marchini and Purwins(2010)]{marchini2010a}
Marco Marchini and Hendrik Purwins.
\newblock Unsupervised generation of percussion sound sequences from a sound
  example.
\newblock In \emph{Sound and Music Computing Conference}, 2010.

\bibitem[Marchini and Purwins(2011)]{marchini2011unsupervised}
Marco Marchini and Hendrik Purwins.
\newblock Unsupervised analysis and generation of audio percussion sequences.
\newblock In \emph{Exploring Music Contents}, pages 205--218. Springer, 2011.

\bibitem[Marxer and Purwins(2010)]{marx10}
Ricard Marxer and Hendrik Purwins.
\newblock Unsupervised incremental learning and prediction of audio signals.
\newblock In \emph{Proc. Int. Symposium on Music Acoustics}, 2010.

\bibitem[Marxer and Purwins(2014)]{marxer14web}
Ricard Marxer and Hendrik Purwins.
\newblock Unsupervised incremental learning and prediction of audio signals:
  Supplementary material, December 2014.
\newblock URL \url{http://www.ricardmarxer.com/research/unsupervised2014}.

\bibitem[Marxer et~al.(2007)Marxer, Holonowicz, Purwins, and Hazan]{marx07}
Ricard Marxer, Piotr Holonowicz, Hendrik Purwins, and Amaury Hazan.
\newblock Dynamical hierarchical self-organization of harmonic, motivic, and
  pitch categories.
\newblock In \emph{Music, Brain and Cognition Workshop, held at Neural
  Information Processing Conference}. 2007.

\bibitem[McKusick and Thompson(1990)]{mcku90}
K.~McKusick and K.~Thompson.
\newblock Cobweb 3: A portable implementation.
\newblock \emph{Technical Report No. FIA-90-6-18-2}, 1990.

\bibitem[Meila(2007)]{meila_comparing_2007}
Marina Meila.
\newblock Comparing clusterings an information based distance.
\newblock \emph{Journal of Multivariate Analysis}, 98\penalty0 (5):\penalty0
  873 -- 895, 2007.

\bibitem[Mermelstein(1976)]{mermelstein1976}
P.~Mermelstein.
\newblock Distance measures for speech recognition, psychological and
  instrumental.
\newblock In \emph{Pattern Recognition and Artificial Intelligence}, pages
  374--388. Academic, New York, 1976.

\bibitem[Mozer(1994)]{mozer1994}
M.C. Mozer.
\newblock Neural network music composition by prediction: Exploring the
  benefits of psychophysical constraints and multiscale processing.
\newblock \emph{Connection Science}, 6:\penalty0 247--280, 1994.

\bibitem[Murphy(2007)]{murphy_conjugate_2007}
Kevin~P. Murphy.
\newblock Conjugate bayesian analysis of the gaussian distribution.
\newblock Technical report, University of British Columbia, 2007.

\bibitem[Pachet(2003)]{pachet2003}
Francois Pachet.
\newblock The continuator: Musical interaction with style.
\newblock \emph{Journal of New Music Research}, 32\penalty0 (3):\penalty0
  333--341, 2003.

\bibitem[Paiement et~al.(2009)Paiement, Grandvalet, and Bengio]{paie2009}
J.F. Paiement, Y.~Grandvalet, and S.~Bengio.
\newblock {Predictive models for music}.
\newblock \emph{Connection Science}, 21\penalty0 (2):\penalty0 253--272, 2009.

\bibitem[Pearce and Wiggins(2004)]{pearce2004}
Marcus~T. Pearce and Geraint~A. Wiggins.
\newblock Improved methods for statistical modelling of monophonic music.
\newblock \emph{Journal of New Music Research}, 33\penalty0 (4):\penalty0
  367--385, 2004.

\bibitem[Pfleger(2002{\natexlab{a}})]{pfle02}
Karl Pfleger.
\newblock \emph{On-line Learning of Predictive Compositional Hierarchies}.
\newblock PhD thesis, Stanford University, 2002{\natexlab{a}}.

\bibitem[Pfleger(2002{\natexlab{b}})]{pfleger02}
Karl Pfleger.
\newblock On-line learning of predictive compositional hierarchies by {H}ebbian
  chunking.
\newblock Technical report, In Proceedings of the AAAI2000 workshop on New
  Research Problems for Machine Learning, 2002{\natexlab{b}}.

\bibitem[Purwins et~al.(2000)Purwins, Blankertz, and
  Obermayer]{purwins_new_2000}
H.~Purwins, B.~Blankertz, and K.~Obermayer.
\newblock A new method for tracking modulations in tonal music in audio data
  format.
\newblock In \emph{Int. Joint Conf. on Neural Network ({IJCNN}'00)}, volume~6,
  pages 270--275. {NI}-{BIT}, 2000.

\bibitem[Ren et~al.(2008)Ren, Dunson, and Carin]{ren_dynamic_2008}
Lu~Ren, David~B Dunson, and Lawrence Carin.
\newblock The dynamic hierarchical dirichlet process.
\newblock In \emph{Proc. Int. Conf. on Machine learning}, pages 824--831.
  {ACM}, 2008.

\bibitem[Smolensky(1986)]{smolensky1986information}
Paul Smolensky.
\newblock Information processing in dynamical systems: Foundations of harmony
  theory.
\newblock In J.~McClelland D.~Rumelhart, editor, \emph{Parallel Distributed
  Processing}, volume~1, pages 194--281. MIT Press, Cambridge, MA, 1986.

\bibitem[Stepleton et~al.(2009)Stepleton, Ghahramani, Gordon, and
  Lee]{stepleton_block_2009}
Thomas~S Stepleton, Zoubin Ghahramani, Geoffrey~J Gordon, and Tai~S Lee.
\newblock The block diagonal infinite hidden {M}arkov model.
\newblock In \emph{Int. Conf. on Artificial Intelligence and Statistics}, pages
  552--559, 2009.

\bibitem[Teh and Jordan(2010)]{teh_hierarchical_2010}
Yee~Whye Teh and Michael~I Jordan.
\newblock Hierarchical {B}ayesian nonparametric models with applications.
\newblock \emph{Journal of the {A}merican Statistical Association}, 1, 2010.

\bibitem[Wagner and Wagner(2007)]{wagner2007comparing}
Silke Wagner and Dorothea Wagner.
\newblock \emph{Comparing clusterings: an overview}.
\newblock Universit{\"a}t Karlsruhe, Fakult{\"a}t f{\"u}r Informatik Karlsruhe,
  2007.

\bibitem[Yoo and Yoo(1995)]{yoo95}
Jungsoon Yoo and Sung Yoo.
\newblock Concept formation in numeric domains.
\newblock In \emph{Proc. ACM Annual Conf. on Computer Science}, pages 36--41,
  1995.

\bibitem[{Zhang} et~al.(2005){Zhang}, {Ghahramani}, and {Yang}]{zhan05}
Jian {Zhang}, Zoubin {Ghahramani}, and Yiming {Yang}.
\newblock A probabilistic model for online document clustering with application
  to novelty detection.
\newblock In \emph{Advances in Neural Information Processing Systems 17}, pages
  1617--1624. MIT Press, Cambridge, MA, 2005.

\bibitem[Zhao and Karypis(2005)]{zhao2005hierarchical}
Y.~Zhao and G~Karypis.
\newblock {Hierarchical clustering algorithms for document datasets}.
\newblock \emph{Data Mining and Knowledge Discovery}, 10\penalty0 (2):\penalty0
  141--168, 2005.

\bibitem[Zitouni et~al.(2002)Zitouni, Siohan, Kuo, and Lee]{zitouni2002backoff}
Imed Zitouni, Olivier Siohan, Hong-Kwang~Jeff Kuo, and Chin-Hui Lee.
\newblock Backoff hierarchical class n-gram language modelling for automatic
  speech recognition systems.
\newblock In \emph{Proc. Interspeech}, 2002.

\end{thebibliography}

%

\newpage
\newpage

\section{Supplement Results: Grid Search on Parameters}

\begin{table}[bh]
\centering
\caption{Cobweb clustering of \emph{Voice} data: ARI (in \%) for different timbral acuities $a$ from Eq.~\ref{eqn:cobweb} (rows) and analysis window lengths  $L$ (Section~\ref{sec:timbre_rep}, columns) .}
\begin{tabular}{@{}c|ccccccccc@{}}\hline
$L \backslash a$
          &        15   &      15.5   &        16   &      16.5   &        17   &      17.5   &        18   &      18.5   &        19   \\\hline
       50 &      31.0   &      33.0   &      33.0   &      33.0   &      33.0   &      34.3   &      35.3   &      35.3   &      35.3   \\
       75 &      62.1   &      57.8   &      42.5   &      42.2   &      46.2   &      46.2   &      34.6   &      34.6   &      36.2   \\
      100 &      76.5   &      77.6   &      75.8   &      78.6   &      81.0   &      78.9   &      55.7   &      55.7   &      37.9   \\
      125 &      79.8   &      79.8   &      71.5   &      73.3   &      73.6   &      73.7   &      78.0   &      78.0   &      80.6   \\
      150 &      67.5   &      66.6   &      67.7   &      68.6   &      73.3   &      76.2   &      81.2   & {\bf 82.7 } &      78.6   \\
      175 &      60.0   &      65.6   &      67.2   &      67.4   &      70.1   &      73.5   &      74.2   &      75.5   &      72.4   \\
\hline
\end{tabular}
\label{tab:clustFrecall}
\end{table}

\begin{table}[bh]
\centering
\caption{Cobweb clustering of \emph{ENST} data: ARI (in \%) for different timbral acuities $a$ from Eq.~\ref{eqn:cobweb} (rows) and analysis window lengths (columns).}
\begin{tabular}{@{}c|ccccc@{}}\hline
$L \backslash a$
          &        13   &      13.5   &        14   &      14.5   &        15   \\\hline
       25 &      65.6   &      64.7   &      63.9   &      63.9   &      62.8   \\
       50 &      83.0   & {\bf 85.7 } &      80.5   &      80.2   &      76.7   \\
       75 &      73.9   &      72.5   &      69.6   &      78.3   &      65.8   \\
      100 &      74.2   &      69.1   &      67.9   &      66.3   &      68.8   \\
\hline
\end{tabular}
\label{tab:clustFmeasureENST}
\end{table}

\begin{table}[bht]
\centering
\caption{Onset and Cobweb transcription: ARI (in \%) of acuity $a$ from Eq.~\ref{eqn:cobweb} for timbre clustering (rows)  versus analysis window length $L$ (columns) measured on the   {\it Voice} data set.}
\begin{tabular}{@{}c|ccccccccc@{}}\hline
$L \backslash a$
          &        15   &      15.5   &        16   &      16.5   &        17   &      17.5   &        18   &      18.5   &        19   \\\hline
       50 &      29.8   &      29.2   &      28.9   &      30.2   &      37.4   &      37.4   &      37.4   &      37.4   &      37.4   \\
       75 &      56.3   &      62.5   &      57.1   &      43.8   &      30.7   &      29.3   &      29.3   &      28.5   &      35.2   \\
      100 &      70.5   &      71.7   &      77.9   &      76.8   &      77.7   &      67.0   &      49.4   &      36.9   &      35.6   \\
      125 &      65.1   &      68.3   &      72.7   &      72.7   &      73.0   &      73.4   &      73.4   &      75.6   &      63.4   \\
      150 &      69.0   &      69.7   &      71.8   &      79.4   & {\bf 81.3 } &      80.7   &      78.8   &      77.7   &      79.0   \\
      175 &      66.4   &      67.7   &      68.4   &      69.8   &      70.8   &      73.4   &      73.4   &      74.6   &      75.6   \\
\hline
\end{tabular}
\label{tab:transVoice}
\end{table}

\begin{table}[bht]
\centering
\caption{Onset and Cobweb transcription: ARI (in \%)  of acuity $a$ from Eq.~\ref{eqn:cobweb} for timbre clustering (rows)  versus analysis window length $L$ (columns) measured on the   {\it ENST} data set.}
\begin{tabular}{@{}c|ccccc@{}}\hline
$L \backslash a$
          &        13   &      13.5   &        14   &      14.5   &        15   \\\hline
       25 &      65.9   &      67.1   &      67.1   &      67.1   &      67.0   \\
       50 &      67.5   & {\bf 76.3 } &      71.4   &      71.4   &      71.4   \\
       75 &      63.2   &      61.4   &      60.6   &      70.8   &      70.2   \\
      100 &      62.4   &      57.2   &      57.9   &      58.1   &      61.6   \\
\hline
\end{tabular}
\label{tab:transENST}
\end{table}

\end{document}